\documentclass[11pt]{article}
\usepackage{hyperref}
\usepackage{amssymb}
\usepackage{amsmath}
\usepackage{mdframed}
\usepackage{subfig}
\usepackage{cancel}
\usepackage{enumerate}
\usepackage{color}
\usepackage{mathtools}
\usepackage{pdfpages}
\usepackage{mathrsfs}
\usepackage{enumitem}
\usepackage{mathtools}
\usepackage{graphicx}
\usepackage{mdframed}
\usepackage{amsfonts}
\usepackage{cancel}
\usepackage{placeins}
\usepackage{amsthm}

\newtheorem{theorem}{Theorem}[section]

\newtheorem{corollary}{Corollary}[section]
\newtheorem{proposition}{Proposition}[section]

\newtheorem{examplex}{Example}[section]
\newenvironment{example}
  {\pushQED{\qed}\examplex}
  {\popQED\endexamplex}

\let\Item\item

\begin{document}

\centerline{{\huge Stochastic approximations of higher-molecular}}

\medskip

\centerline{{\huge by bi-molecular reactions}}
 
\medskip
\bigskip

\centerline{
\renewcommand{\thefootnote}{$*$}
{\Large Tomislav Plesa\footnote{
Department of Bioengineering, Imperial College London,
Exhibition Road, \\ London, SW7 2AZ, UK;
e-mail: t.plesa@ic.ac.uk}}}

\medskip
\bigskip

\noindent
{\bf Abstract}:
Biochemical reactions involving three or more reactants, called higher-molecular reactions,
play an important role in theoretical systems and synthetic biology. In particular, 
such reactions underpin a variety of important bio-dynamical phenomena, such as
multi-stability/multi-modality, oscillations, bifurcations, and noise-induced effects. 
However, only reactions with at most two reactants, called bi-molecular reactions, 
are experimentally feasible. To bridge the gap, in this paper we put forward an algorithm 
for systematically approximating arbitrary higher-molecular reactions with bi-molecular ones, 
while preserving the underlying stochastic dynamics. Properties of the algorithm and 
convergence are established via singular perturbation theory. 
The algorithm is applied to a variety of higher-molecular biochemical networks, 
and is shown to play an important role in nucleic-acid-based synthetic biology.


\section{Introduction}
Reaction networks~\cite{Feinberg,Janos} are a central mathematical framework
for analyzing biochemical processes from systems biology~\cite{SysBio1,SysBio2,SysBio3},
and are a powerful programming language for designing 
molecular systems in synthetic biology~\cite{DNAComputing1,DNAComputing2,MeRobust,Me1,Me3,DNAComputing3}. 
Experimentally feasible biochemical networks contain only reactions with at most
two reacting molecules (second-order/bi-molecular reactions), as
reactive collisions between three or more molecules (higher-order/higher-molecular reactions)
are unlikely to take place~\cite{Janos,Gillespietri}. 
For example, in nucleic-acid-based synthetic biology, where abstract biochemical networks
are experimentally implemented with nucleic acids (DNA or RNA molecules)~\cite{Experiment5}, 
only second-order reactions have been rigorously shown to be realizable~\cite{DNAComputing1}.
Despite experimental implausibility, higher-order reactions 
appear in both theoretical systems and synthetic biology.
For example, third-order (tri-molecular) reactions appear 
in the one-species Schl\"{o}gl system~\cite{Schlogl}, where they allow
for bi-stability (coexistence of two stable equilibria), 
in the Brusselator~\cite{Brusselator} 
and Schnakenberg systems~\cite{Schnakenberg}, 
which display oscillations (existence of a stable limit cycle), 
as well as in two-species biochemical networks displaying
bicyclicity (coexistence of two stable limit cycles)~\cite{Me2}, 
and homoclinic and SNIC bifurcations~\cite{Me1,RadekSNIPER}.
Aside from well-mixed settings, third-order
reactions also play a role in pattern formation~\cite{Radek2}
and, more broadly, are a subject of research within reaction-diffusion
bio-modelling~\cite{Radek3}.
In context of synthetic biology, higher-order reactions 
appear in the noise-control algorithm~\cite{Me3}
and the stochastic morpher controller~\cite{MeRobust}, 
where they allow for local and global reshaping
of the probability distributions of the molecular species, respectively. 

An algorithm for approximating third- and fourth-order reactions with second-order ones
at the \emph{deterministic level}, i.e. at the level of the reaction-rate equations~\cite{Feinberg},
has been used for decades~\cite{Highmol1,Highmol2,UNI3}.
The algorithm relies on suitable time-scale separations, and
has been formally justified for third- and fourth-order reactions
at the deterministic level~\cite{Highmol2,UNI3}.
Another, more elaborate, order-reduction procedure has been presented in~\cite{UNI1,Kowalski};
while this procedure does not rely on time-scale separations,
it depends on the precise initial conditions of some of the underlying species
which, from the perspective of synthetic biology, may pose
significant challenges~\cite{Vesicles2,Vesicles3}. 
Less attention has been paid to the validity of such approximations at the \emph{stochastic level},
i.e. at the level of the chemical master equation (CME)~\cite{GillespieDerivation}.
In this context, it has been formally shown in~\cite{Janssen} that a specific third-order reaction,
namely $3 X \to 2 X$, can be stochastically approximated with a second-order network using the algorithm 
from~\cite{Highmol1,Highmol2,UNI3}, and this has also been qualitatively described in~\cite{Gillespietri}. 
However, the questions of convergence and whether the formal 
deterministic results from~\cite{Highmol1,Highmol2,UNI3}
extend into the stochastic regime for arbitrary reactions remain unanswered.
In particular, validity of perturbation results at the deterministic level does not generally 
imply validity at the stochastic level~\cite{GrimaQSA,KimQSA,AgarwalQSA}.
In this paper, we generalize the algorithm from~\cite{Highmol1,Highmol2,UNI3}
to stochastic biochemical networks with arbitrary number and composition of reactants,
and we provide convergence analyses. 

The paper is organized as follows. 
In Section~\ref{sec:trimolecular},
we prove that any one-species third-order reaction
can be approximated with a suitable family
of second-order networks, and we apply 
the results in Section~\ref{sec:Schlogl} 
on the Schl\"{o}gl system~\cite{Schlogl}.
In Section~\ref{sec:nmolecular}, we generalize
the results from Section~\ref{sec:trimolecular} to 
arbitrary multi-species higher-order reactions
under mass-action kinetics. 
In Section~\ref{sec:examples}, 
we apply the generalized results to higher-order
reaction networks displaying noise-induced phenomena.
Finally, we conclude with a summary and discussion in
 Section~\ref{sec:discussion}. The notation and background 
theory used in the paper are introduced as needed, and are summarized 
in Appendix~\ref{app:background}. Detailed analyses 
supporting the results from Section~\ref{sec:nmolecular} 
are provided in Appendices~\ref{app:formal}--\ref{app:convergence}. 

\section{Special case: One-species third-order reactions} \label{sec:trimolecular}
Let us consider an arbitrary one-species third-order (tri-molecular) reaction,
under mass-action kinetics~\cite{Feinberg}, given by
\begin{align}
\mathcal{R}_0(X) : \; \; 3 X  &\xrightarrow[]{k} \bar{\nu} X, \label{eq:reduced3}
\end{align}
where $X$ is a biochemical species, $\bar{\nu} \in \mathbb{Z}_{\ge}$ is a 
product stoichiometric coefficients, and $k \in \mathbb{R}_{>}$ is a dimensionless rate coefficient; 
here, $\mathbb{Z}_{\ge}$ and $\mathbb{R}_{>}$ are the sets of nonnegative
integers and positive real numbers, respectively, see also Appendix~\ref{app:background} 
for notation and reaction network theory.
Consider also the second-order (bi-molecular) mass-action network
$\mathcal{R}_{\varepsilon}(X, Y) = \mathcal{R}_{0}^{\varepsilon}
\cup  \mathcal{R}_{1} \cup \mathcal{R}_{2}$, given by
\begin{align}
\mathcal{R}_{0}^{\varepsilon}(X, Y): \hspace{0.3cm}
&  Y  \xrightarrow[]{1/\varepsilon} 2 X, \nonumber \\
\mathcal{R}_{1}(X, Y): \hspace{0.3cm}
& 2 X   \xrightarrow[]{\kappa_1}  Y, \nonumber \\
\mathcal{R}_{2}(X, Y): \hspace{0.3cm}
&  X + Y   \xrightarrow[]{\kappa_2} \tilde{\nu} X  + \bar{\gamma} Y,
\label{eq:subnetworks3}
\end{align}
where $Y$ is an auxiliary species, $\tilde{\nu}, \bar{\gamma} \in \mathbb{Z}_{\ge}$, and 
$\kappa_1, \kappa_2 \in \mathbb{R}_{>}$ are dimensionless rate coefficients.
Network~(\ref{eq:subnetworks3}) is said to be of second-order because
its highest-order reaction is of second-order.
In what follows, we prove that, under suitable conditions on the 
kinetic and stoichiometric coefficients $\kappa_1, \kappa_2$ and 
$\tilde{\nu}, \bar{\gamma}$, respectively, the $x$-marginal probability-mass function (PMF) 
of reaction network~(\ref{eq:subnetworks3}) approaches the PMF of~(\ref{eq:reduced3}) 
as $\varepsilon \to 0$, which we formulate as 
Proposition~\ref{proposition:convergence3} in Section~\ref{sec:conv3}.

\subsection{Perturbation analysis} \label{sec:perturbation3}
Let us denote the discrete copy-numbers of species 
$\{X, Y\}$ by $(x, y) \in \mathbb{Z}_{\ge}^{2}$, 
and the continuous time-variable by $t \in \mathbb{R}_{\ge}$.
Under suitable conditions, PMF of reaction network~(\ref{eq:subnetworks3}), 
denoted by $p_{\varepsilon}(x,y,t)$, satisfies a partial difference-differential
equation, called the \emph{chemical master equation} (CME)~\cite{GillespieDerivation,RadekBook,VanKampen}, 
see also Appendix~\ref{app:background}.
As motivated shortly, we introduce new coordinates 
$\bar{x} =  (x + 2 y)$ and $\tau = \varepsilon t$, in which the CME for~(\ref{eq:subnetworks3}) reads
\begin{align}
\frac{\mathrm{d}}{\mathrm{d} \tau} p_{\varepsilon}(\bar{x},y,\tau) & 
= \left(\frac{1}{\varepsilon^{2}}\mathcal{L}_0  + \frac{1}{\varepsilon} (\mathcal{L}_1 + \mathcal{L}_2) \right)  
p_{\varepsilon}(\bar{x},y,\tau), 
\; \; \textrm{where } \bar{x} =  x + 2 y,
\label{eq:CME3}
\end{align}
where operators $\mathcal{L}_0$, $\mathcal{L}_1$ and $\mathcal{L}_2$
are induced by reactions $\mathcal{R}_{0}^1$, $\mathcal{R}_1$ and $\mathcal{R}_2$, 
respectively, and read
\begin{align}
\mathcal{L}_0  & = \left(E_{y}^{+1} - 1 \right) y, \nonumber \\  
\mathcal{L}_{1}  & = \left(E_{y}^{-1} - 1 \right) \alpha_1(\bar{x},y), 
\; \; \hspace{2.9cm} \textrm{where } \alpha_1(\bar{x},y) = \bar{\kappa}_1 (\bar{x} - 2 y) (\bar{x} - 2 y - 1), \nonumber \\
\mathcal{L}_{2}  & = \left(E_{\bar{x}}^{-(\tilde{\nu} + 2 \bar{\gamma} - 3)} E_{y}^{- (\bar{\gamma} - 1)} - 1 \right) \alpha_{2}(\bar{x},y) y, 
\; \; \; \textrm{where } \alpha_2(\bar{x},y) = \bar{\kappa}_2 (\bar{x} - 2 y),
\label{eq:operators3}
\end{align}
with a step operator $E_{x}^{-\Delta x}$ such that 
$E_{x}^{-\Delta x} p(x,t) = p(x - \Delta x,t)$.

Let us consider the perturbation series
\begin{align}
p_{\varepsilon}(\bar{x},y,\tau) & = p_0(\bar{x},y,\tau) +  \varepsilon p_1(\bar{x},y,\tau)  
+ \varepsilon^2 p_2(\bar{x},y,\tau) + \ldots, \label{eq:perturbation3}
\end{align} 
where we require that the zero-order term is a PMF, i.e. 
$p_0(\cdot, \cdot,\tau) : \mathbb{Z}_{\ge}^{2} \to [0,1]$ and 
$\langle 1,  p_0(\bar{x},y,\tau) \rangle$
$\equiv \sum_{\bar{x}} \sum_{y} p_0(\bar{x},y,\tau) = 1$ for all $\tau \ge 0$. 
Substituting~(\ref{eq:perturbation3}) into~(\ref{eq:CME3}), differentiating 
the series term-wise with respect to time, and equating terms of equal powers in $\varepsilon$, 
one obtains the following system of equations:
\begin{align}
\mathcal{L}_{0} p_0(\bar{x},y,\tau) & = 0, \label{eq:QSA3a} \\
\mathcal{L}_{0} p_{1}(\bar{x},y,\tau) & = 
- (\mathcal{L}_1 + \mathcal{L}_2) p_{0}(\bar{x},y,\tau), \label{eq:QSA3b} \\
\mathcal{L}_{0} p_{2}(\bar{x},y,\tau) & = 
\frac{\mathrm{d}}{\mathrm{d} \tau} p_{0}(\bar{x},y,\tau)
- (\mathcal{L}_1 + \mathcal{L}_2) p_{1}(\bar{x},y,\tau). \label{eq:QSA3c}
\end{align}

\emph{Equation}~(\ref{eq:QSA3a}). 
Since operator $\mathcal{L}_0$ acts and depends only on $y$, 
we seek the zero-order term in a separable form,
 $p_0(\bar{x},y,\tau) = p_0(\bar{x},\tau) p_0(y)$, so that
\begin{align}
p_0(\bar{x},y,\tau) & = p_0(\bar{x},\tau) \delta_{y, 0},
\label{eq:sol3p0}
\end{align}
where $\delta_{y,0}$ is the Kronecker-delta distribution centered
at zero, see also Appendix~\ref{app:background}. 

\noindent \emph{Remark}. In the original coordinates $(x, y)$, 
operator $\mathcal{L}_0$ induces reaction $Y \xrightarrow[]{1} 2 X$, 
and~(\ref{eq:QSA3a}) has infinitely many solutions. 
This degeneracy arises from the fact that
the process induced by $\mathcal{L}_0$ satisfies a local linear
conservation law $x + 2 y = \bar{x}$, 
where $\bar{x}$ is time-independent.
Using this conservation law as a coordinate change
ensures that~(\ref{eq:QSA3a}) has a unique solution 
$p_0(y) = \delta_{y,0}$.

\emph{Equation}~(\ref{eq:QSA3b}). 
Using~(\ref{eq:operators3}) and~(\ref{eq:sol3p0}), it follows that
\begin{align}
(\mathcal{L}_1 + \mathcal{L}_2) p_0(\bar{x},y,\tau) & = 
\mathcal{L}_1 p_0(\bar{x},y,\tau) =
p_0(\bar{x},\tau) \left(E_{y}^{-1} - 1 \right) 
\alpha_1(\bar{x},y) \delta_{y, 0}. \label{eq:solvabilitycond3}
\end{align}
Since the null-space of $\mathcal{L}_0$ is non-trivial, 
equation~(\ref{eq:QSA3b}) has either no solutions or infinitely many solutions. 
Let $\langle f, g\rangle_{y} \equiv \sum_{y = 0}^{\infty} f(y) g(y)$ 
denote an $l^2$ inner-product, and note that $\mathcal{L}_0^* = 
y \left(E_{y}^{-1} - 1 \right)$ is the $l^2$-adjoint 
(backward) operator corresponding to $\mathcal{L}_0$, see also Appendix~\ref{app:background}. 
The null-space of  $\mathcal{L}_0^*$ is given by constants in $y$, 
$\mathcal{N}(\mathcal{L}_0^*) = \{1\}$. Therefore, since 
$\langle 1, (\mathcal{L}_1 + \mathcal{L}_2) p_0(\bar{x},y,\tau) \rangle_{y}
= \langle \mathcal{L}_1^* 1, p_0(\bar{x},y,\tau) \rangle_{y} = 0$,
the Fredholm alternative theorem implies that~(\ref{eq:QSA3b}) has a solution. 

\noindent \emph{Remark}. In the original coordinate $t$, 
the Fredholm alternative theorem enforces
a trivial effective CME $\mathrm{d}/\mathrm{d} t p_0(\bar{x},y,t) = 0$; 
to capture non-trivial dynamics, we have rescaled time 
to a longer scale. 

Considering the form of $\mathcal{L}_0$ and~(\ref{eq:solvabilitycond3}), 
we seek a solution of~(\ref{eq:QSA3b}) in a separable form 
\begin{align}
p_1(\bar{x},y,\tau)  & = p_0(\bar{x},\tau) p_1(y; \, \bar{x}). \label{eq:p13}
\end{align}
Substituting~(\ref{eq:solvabilitycond3})--(\ref{eq:p13}) into~(\ref{eq:QSA3b}), one obtains
\begin{align}
y p_1(y; \, \bar{x}) & = E_{y}^{-1} 
\alpha_1(\bar{x},y) \delta_{y, 0}. \label{eq:so3lp1}
\end{align}

\emph{Equation}~(\ref{eq:QSA3c}).
Applying the Fredholm alternative theorem to~(\ref{eq:QSA3c}), 
and using~(\ref{eq:operators3}), (\ref{eq:p13})
and~(\ref{eq:so3lp1}), one obtains
\begin{align}
 \frac{\mathrm{d}}{\mathrm{d} \tau} p_0(\bar{x},\tau) 
& =  \left \langle 1, \mathcal{L}_{2} p_{1}(\bar{x},y,\tau) \right \rangle_{y}
= (E_{\bar{x}}^{-\Delta \bar{x}} - 1) \alpha_1(\bar{x},0)  \alpha_{2}(\bar{x},1) 
p_{0}(\bar{x},\tau).
\label{eq:solvability3}
\end{align}
Substituting~(\ref{eq:operators3})  into~(\ref{eq:solvability3}), 
and using $\tau = \varepsilon t$, 
one obtains the \emph{effective} CME 
\begin{align}
 \frac{\mathrm{d}}{\mathrm{d} t}  p_0(\bar{x},t)  & = 
\left(E_{\bar{x}}^{-(\tilde{\nu} + 2 \bar{\gamma} - 3)} - 1 \right) 
\varepsilon \kappa_1 \kappa_2 \bar{x} (\bar{x} - 1) (\bar{x} - 2) p_0(\bar{x},t). 
\label{eq:effectiveCME3}
\end{align}

\noindent \emph{Remark}. Equation~(\ref{eq:effectiveCME3}) describes a time-evolution
of the PMF for the stochastic process $\bar{X}(t) = X(t) + 2 Y(t)$, and not
the original copy-numbers $X(t)$. However, equation~(\ref{eq:sol3p0})
implies that process $Y(t)$ spends most of the time at $y = 0$
as $\varepsilon \to 0$, so that PMFs for $\bar{X}(t)$ 
and $X(t)$ match as $\varepsilon \to 0$. 

\subsection*{Kinetic and stoichiometric conditions}
In order to ensure that the dynamics of networks~(\ref{eq:reduced3}) and~(\ref{eq:subnetworks3})
match, coefficients $\kappa_1, \kappa_1$ have to suitably scale with $\varepsilon$.
In particular, the CME for network~(\ref{eq:reduced3}) is given by
\begin{align}
 \frac{\mathrm{d}}{\mathrm{d} t}  p(x,t)  & = 
\left(E_{x}^{-(\bar{\nu} - 3)} - 1 \right) 
k x (x - 1) (x - 2) p(x,t). \label{eq:inputCME3}
\end{align}
In order for~(\ref{eq:effectiveCME3}) and~(\ref{eq:inputCME3}) 
to match, we impose the \emph{kinetic condition}, given by
\begin{align}
\varepsilon \kappa_1 \kappa_2  & =  k, \; \; \textrm{where }
\kappa_1, \kappa_2 = o(\varepsilon^{-1}) \; \; \text{as } \varepsilon \to 0,
\label{eq:kinetic3}
\end{align}
and the \emph{stoichiometric condition}, given by
\begin{align}
\tilde{\nu} & =  \bar{\nu} - 2 \bar{\gamma}, \label{eq:stoichiometric3}
\end{align}
where $o(\cdot)$ is the ``little-o" asymptotic symbol, see also Appendix~\ref{app:background}.

\noindent \emph{Remark}. Requirement $\kappa_1, \kappa_2 = o(\varepsilon^{-1})$
as $\varepsilon \to 0$ from~(\ref{eq:kinetic3}) ensures that that 
operator $\mathcal{L}_0$ from~(\ref{eq:operators3}) does not change; 
otherwise, if $\kappa_1 = \mathcal{O}(\varepsilon^{-1})$ or 
$\kappa_2 = \mathcal{O}(\varepsilon^{-1})$, where 
$O(\cdot)$ is the ``big-O" symbol, one obtains families of perturbation problems 
distinct from the one considered in this section. 

\subsection{Convergence} \label{sec:conv3}
The formal perturbation analysis from Section~\ref{sec:perturbation3} has been 
performed under the assumption that $\kappa_1$ and $\kappa_2$ are independent 
of $\varepsilon$, which is inconsistent with the kinetic condition~(\ref{eq:kinetic3}).
We stress that an objective of the analysis in Section~\ref{sec:perturbation3} 
was precisely to uncover admissible $\varepsilon$-scaling of $\kappa_1$ and $\kappa_2$
ensuring that~(\ref{eq:reduced3}) and~(\ref{eq:subnetworks3}) match. 
Having formally obtained such candidates, we now perform a convergence analysis. 
In particular, let us satisfy~(\ref{eq:kinetic3}) with
\begin{align}
\kappa_1 &= \bar{\kappa}_1 \varepsilon^{-1/2}, \; \; \; 
 \kappa_2 = \bar{\kappa}_2 \varepsilon^{-1/2}, \label{eq:scaling3}
\end{align}
where $\bar{\kappa}_1, \bar{\kappa}_2$ are $\varepsilon$-independent parameters.
The CME for network~(\ref{eq:subnetworks3}) under~(\ref{eq:scaling3}) is given by
\begin{align}
\frac{\mathrm{d}}{\mathrm{d} \tau} p_{\varepsilon}(\bar{x},y,t) & 
= \left(\frac{1}{\varepsilon^{2}}\mathcal{L}_0  + 
\frac{1}{\varepsilon^{1/2}} (\mathcal{L}_1 + \mathcal{L}_2) \right)  
p_{\varepsilon}(\bar{x},y,t), 
\; \; \textrm{where } \bar{x} =  x + 2 y.
\label{eq:CME3_scaled}
\end{align}
Substituting into~(\ref{eq:CME3_scaled}) the fractional-power perturbation series 
\begin{align}
p_{\varepsilon}(\bar{x},y,t)  & =  p_0(\bar{x},y,t) + \varepsilon^{1/2} p_1(\bar{x},y,t) + \varepsilon p_2(\bar{x},y,t)  + \ldots,
\label{eq:perturbation3_scaled}
\end{align} 
one obtains
\begin{align}
\mathcal{O} \left(\frac{1}{\varepsilon} \right): \;  
\mathcal{L}_{0} p_0(\bar{x},y,t) & = 0, \nonumber \\
\mathcal{O} \left(\frac{1}{\varepsilon^{1/2}} \right): \;  
\mathcal{L}_{0} p_{1}(\bar{x},y,t) & = 
- (\mathcal{L}_1 + \mathcal{L}_2) p_{0}(\bar{x},y,t), \nonumber \\
\mathcal{O} \left(1 \right): \;  
\mathcal{L}_{0} p_{2}(\bar{x},y,t) & = 
\frac{\mathrm{d}}{\mathrm{d} \tau} p_{0}(\bar{x},y,t)
- (\mathcal{L}_1 + \mathcal{L}_2) p_{1}(\bar{x},y,t). \label{eq:QSA3_scaled}
\end{align}
System~(\ref{eq:QSA3_scaled}) has the same solutions 
as~(\ref{eq:QSA3a})--(\ref{eq:QSA3c}), since the two system have
identical forms. In particular, the zero-order PMF is given by
\begin{align}
p_0(\bar{x},y,t) & = p_0(\bar{x},t) \delta_{y, 0},
\label{eq:sol3p0_scaled}
\end{align}
where the factor $p_0(\bar{x},t)$ satisfies
\begin{align}
 \frac{\mathrm{d}}{\mathrm{d} t}  p_0(\bar{x},t)  & = 
\left(E_{\bar{x}}^{-(\tilde{\nu} + 2 \bar{\gamma} - 3)} - 1 \right) 
\bar{\kappa}_1 \bar{\kappa}_2 \bar{x} (\bar{x} - 1) (\bar{x} - 2) p_0(\bar{x},t). 
\label{eq:effectiveCME3_scaled}
\end{align}

In what follows, we establish a weak convergence result 
over bounded domains; to this end, we denote the $l_1$-norm over a bounded
set $\mathbb{S}$ by $\| \cdot \|_{l_1(\mathbb{S})}$.
\begin{proposition} \label{proposition:convergence3} 
\textit{Consider network $\mathcal{R}_{\varepsilon}$, 
given by~{\rm(\ref{eq:subnetworks3})}, with the rate coefficients
$\kappa_1$ and $\kappa_2$ given by~{\rm(\ref{eq:scaling3})},
and with the {\rm PMF} $p_{\varepsilon}(\bar{x},y,t)$ satisfying {\rm(\ref{eq:CME3_scaled})}.
Let the {\rm PMF} $p_0(\bar{x},y,t)$ 
satisfy~{\rm (\ref{eq:sol3p0_scaled})}--{\rm (\ref{eq:effectiveCME3_scaled})}, 
and assume that $p_{\varepsilon}(\bar{x},y,0) = p_0(\bar{x},y,0)$.
Then, for every $\mathbb{S} \subset \mathbb{Z}_{\ge}^{2}$
and every $T > 0$ there exists an $\varepsilon$-independent constant $c > 0$ such that
\begin{align}
 \left \|p_{\varepsilon}(\bar{x},y,t) - p_0(\bar{x},y,t) \right \|_{l_1(\mathbb{S})} & 
\le  c \, \varepsilon^{1/2}, \; \;  \textrm{as } \varepsilon \to 0,
\; \; \textrm{for all } t \in [0,T].
\label{eq:convergence3}
\end{align}
}
\end{proposition} 

\begin{proof}
By construction, there exist functions $p_1(\bar{x},y,t)$ and $p_2(\bar{x},y,t)$
such that system~(\ref{eq:QSA3_scaled}) is satisfied; in what follows, we write 
$p_i(t) = p_i(\bar{x},y,t)$ for all $i \in \{0, 1, 2\}$.
Let us define a remainder $r_{\varepsilon}(t) = r_{\varepsilon}(\bar{x},y,t)$ via
\begin{equation}
p_{\varepsilon}(t) = p_0(t) + \varepsilon^{1/2} p_1(t) + \varepsilon p_2(t) + r_{\varepsilon}(t).
 \label{eq:remainder3}
\end{equation}
Substituting~(\ref{eq:remainder3}) into~(\ref{eq:CME3_scaled}), using~(\ref{eq:QSA3_scaled}) 
and the assumption that $p_{\varepsilon}(0) = p_0(0)$, 
one obtains an initial-value problem for the remainder:
\begin{align}
\frac{\mathrm{d}}{\mathrm{d} t} r_{\varepsilon} (t) - \mathcal{L}_{\varepsilon} r_{\varepsilon} (t) & = 
\varepsilon^{\frac{1}{2}} \left(- \frac{\mathrm{d}}{\mathrm{d} t} p_1(t) + (\mathcal{L}_1 + \mathcal{L}_2) p_2(t) \right) 
- \varepsilon \frac{\mathrm{d}}{\mathrm{d} t} p_2(t),
\; \; \; 
r_{\varepsilon}(0) = -\left(\varepsilon^{\frac{1}{2}} p_1(0) + \varepsilon p_2(0) \right).
\label{eq:error3}
\end{align}
Solving~(\ref{eq:error3}), applying the $l^1$-norm on a bounded set 
$\mathbb{S} \subset \mathbb{Z}_{\ge}^{2}$, the triangle inequality, 
and using the fact that $\| e^{\mathcal{L}_{\varepsilon} t} \|_{l_1(\mathbb{S})} \le 1$, one obtains
\begin{align}
 \| r_{\varepsilon} (t) \|_{l_1(\mathbb{S})} & \le  
\varepsilon^{\frac{1}{2}} \left[ \| p_1(0)\|_{l_1(\mathbb{S})}
+ t \, \textrm{sup}_{0 \le s \le t} \left( \left \| \frac{\mathrm{d}}{\mathrm{d} s} p_1(s) \right\|_{l_1(\mathbb{S})} 
+ \left(\|\mathcal{L}_1 \|_{l_1(\mathbb{S})}  + \|\mathcal{L}_2\|_{l_1(\mathbb{S})} \right) 
\|p_2(s)\|_{l_1(\mathbb{S})} \right) \right]
\nonumber \\
& +
\varepsilon \left[\| p_2(0)\|_{l_1(\mathbb{S})}
+ t \, \textrm{sup}_{0 \le s \le t} \left \|\frac{\mathrm{d}}{\mathrm{d} s} p_2(s) \right\|_{l_1(\mathbb{S})} \right].
\label{eq:errorbound3}
\end{align}
It follows from~(\ref{eq:QSA3_scaled}) that there exist
$p_1(t)$ and $p_2(t)$ that are bounded with bounded time-derivatives; 
therefore, $\| r_{\varepsilon} (t) \|_{l_1(\mathbb{S})} = \mathcal{O}(\varepsilon^{1/2})$
as $\varepsilon \to 0$ for all $t \in [0,T]$ which, together 
with~(\ref{eq:remainder3}), implies~(\ref{eq:convergence3}).
\end{proof}
\noindent \emph{Remark}. The assumption $p_{\varepsilon}(\bar{x},y,0) = p_0(\bar{x},y,0)$
can be removed from Proposition~\ref{proposition:convergence3} under a suitable 
initial-layer analysis, which we do not pursue in this paper. 

\noindent \emph{Remark}. Error estimate~(\ref{eq:convergence3}) is
independent of the stoichiometric coefficients $\tilde{\nu}$ and $\bar{\gamma}$. 

Generalizing the analysis from this section 
(see also Appendix~\ref{app:formal}), one can show that the 
error under a fractional-power scaling
$\kappa_1 = \bar{\kappa}_1 \varepsilon^{-n/d}$ and 
$\kappa_2 = \bar{\kappa}_2 \varepsilon^{-(1 - n/d)}$, 
with $n, d \in \mathbb{Z}_{>}$ and $n/d < 1$, is asymptotically 
bounded by $c \varepsilon^{1/d}$ for some $c > 0$. 

\section{Example: The Schl\"{o}gl network} \label{sec:Schlogl}
\begin{figure}[!htbp]
\vskip  0.2cm
\centerline{
\includegraphics[width=0.4\columnwidth]{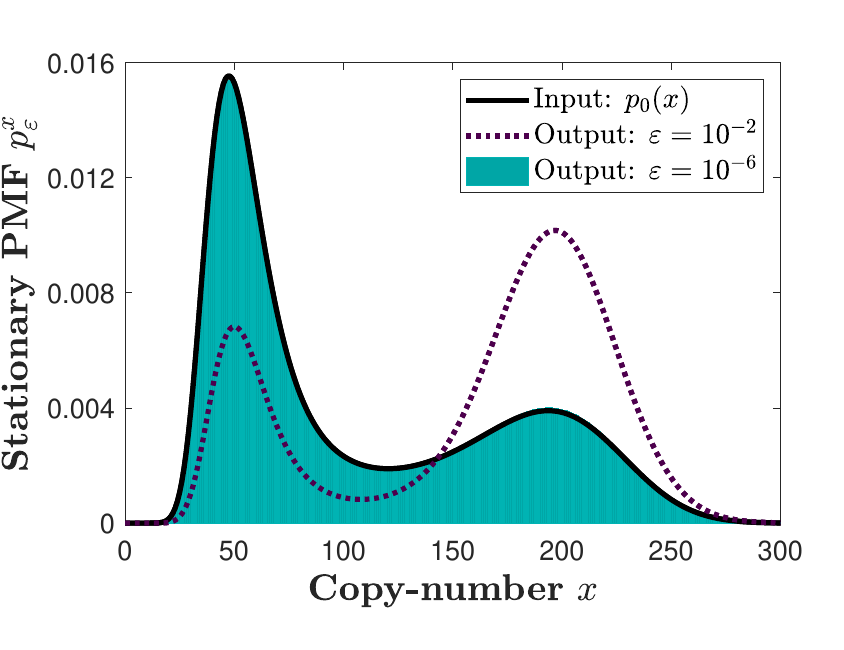}
\hskip 1mm
\includegraphics[width=0.4\columnwidth]{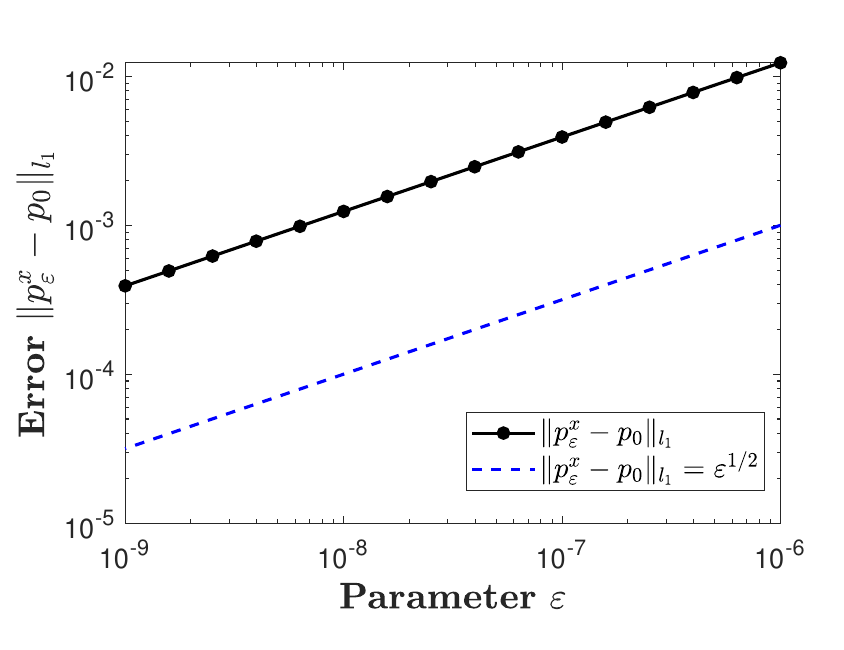}
}
\vskip -5.4cm
\leftline{\hskip 1.8cm (a) \hskip 6.4cm (b)}
\vskip 5.3cm
\centerline{
\includegraphics[width=0.4\columnwidth]{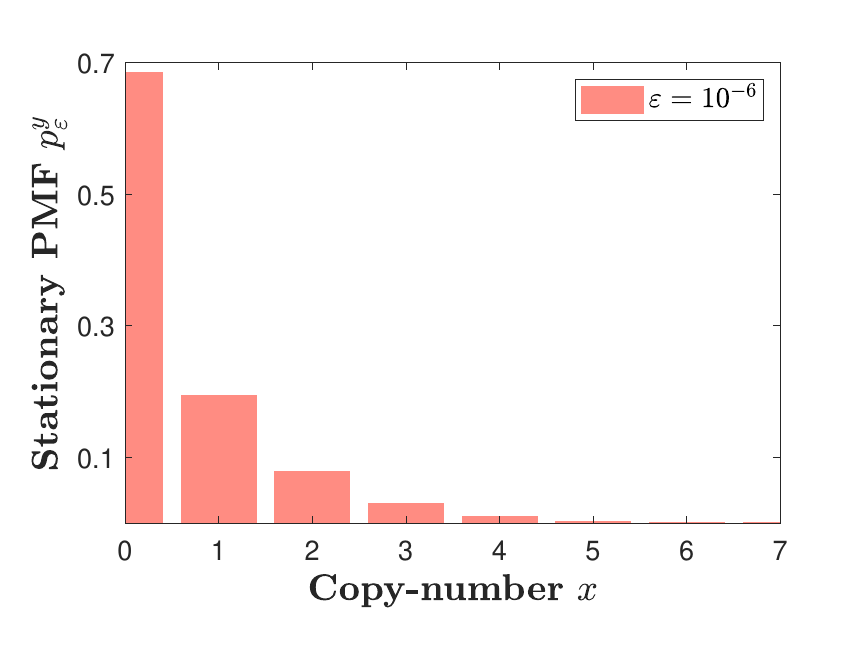}
\hskip 1mm
\includegraphics[width=0.4\columnwidth]{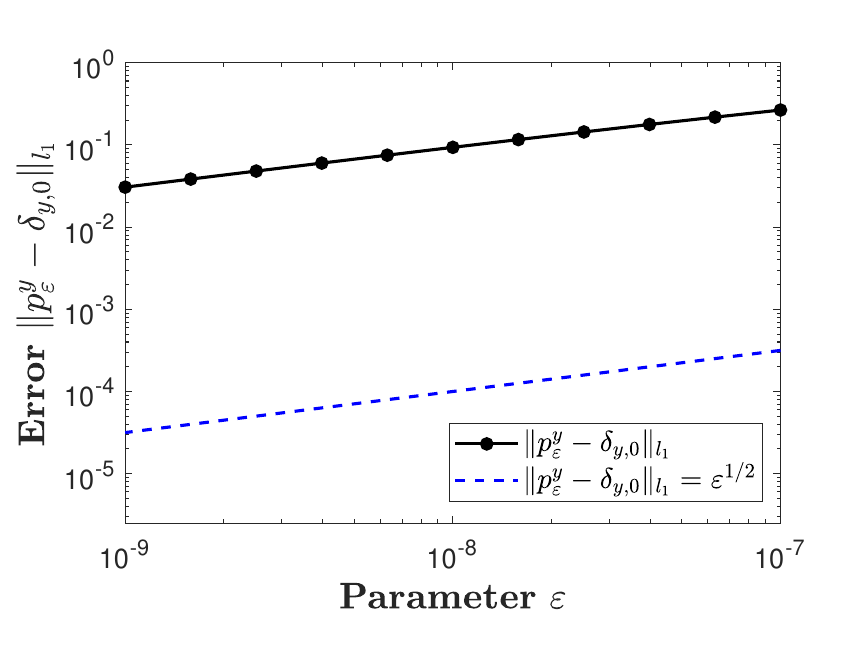}
}
\vskip -5.4cm
\leftline{\hskip 1.8cm (c) \hskip 6.4cm (d)}
\vskip 4.4cm
\caption{\it{Panel {\rm (a)} 
displays the stationary {\rm PMF} of the input network~{\rm (\ref{eq:netschlogl})}
as a black curve, and the $x$-marginal {\rm PMF} for the output network~
{\rm (\ref{eq:netschloglmodified})}, under~{\rm(\ref{eq:scaling3_ex})}, 
for different values of the asymptotic parameter $\varepsilon$.
Panel {\rm (b)} displays as black dots, interpolated with the solid lines, a
 log-log plot of the $l^1$-distance between the {\rm PMF}s 
for networks~{\rm (\ref{eq:netschlogl})} and~{\rm (\ref{eq:netschloglmodified})}
as a function of $\varepsilon$. Analogous plots are shown in 
panels {\rm (c)} and {\rm (d)} for the $y$-marginal {\rm PMF} for~{\rm (\ref{eq:netschloglmodified})}.
 The parameters are fixed to $(k_1,k_2,k_3,k_4) = (1125,37.5,0.36,10^{-3})$,
 and $(\tilde{\nu},\bar{\gamma}) = (0,1)$.}} \label{fig:Schlogl1}
\end{figure}

In this section, we apply the results developed in Section~\ref{sec:trimolecular} 
to the one-species third-order Schl\"{o}gl network~\cite{Schlogl}, given by
\begin{align}
 \mathcal{R}_0(X): \; \; \; \varnothing  &
\xrightleftharpoons[k_2]{k_1} X, \nonumber \\
  2 X &
\xrightleftharpoons[k_4]{k_3} 3 X.
  \label{eq:netschlogl}
\end{align} 
Here, $\varnothing$ represents species that are not explicitly modelled, 
and the irreversible forward and backward reactions $\varnothing \xrightarrow[]{k_1} X$
and $X \xrightarrow[]{k_2} \varnothing$, respectively,
are jointly denoted as a single reversible reaction 
$\varnothing \xrightleftharpoons[k_2]{k_1} X$, and similarly
for $  2 X \xrightleftharpoons[k_4]{k_3} 3 X$.
In Figure~\ref{fig:Schlogl1}(a), for a particular choice of the rate coefficients, 
we display as a black curve the stationary PMF for the input network~(\ref{eq:netschlogl}), 
denoted by $p_0 = p_0(x)$, which displays two maxima (bi-modality).
Approximating the third-order reaction from~(\ref{eq:netschlogl}) 
with~(\ref{eq:subnetworks3}), one obtains
\begin{align}
\mathcal{R}_{\varepsilon}(X, Y): \; \; \; \; \; \; \varnothing  &
\xrightleftharpoons[k_2]{k_1} X, \nonumber \\
 2 X & \xrightarrow[]{ k_3 } 3 X, \nonumber \\
2 X & \xrightleftharpoons[1/\varepsilon]{\bar{\kappa}_1 \varepsilon^{-s}} Y, \nonumber \\
X + Y & \xrightarrow[]{\bar{\kappa}_2 \varepsilon^{-(1 - s)}} 
\tilde{\nu} X + \bar{\gamma} Y, 
\; \; \; \; \textrm{where } s \in (0,1).\label{eq:netschloglmodified}
\end{align}
\newpage
The stoichiometric condition~(\ref{eq:stoichiometric3})
demands that $\tilde{\nu} = (2 - 2 \bar{\gamma})$, and there are two
choices: taking $\bar{\gamma} = 0$ implies that $\tilde{\nu} = 2$,
taking $\bar{\gamma} = 1$ implies that $\tilde{\nu} = 0$, 
while taking $\bar{\gamma} \ge 2$ implies that $\tilde{\nu} < 0$, 
which is biochemically infeasible. In what follows, we
take $(\tilde{\nu},\bar{\gamma}) = (0,1)$, and consider 
different scaling factors $s$ to satisfy~(\ref{eq:kinetic3}).

\begin{figure}[!htbp]
\vskip  0.2cm
\centerline{
\hskip 1mm
\includegraphics[width=0.4\columnwidth]{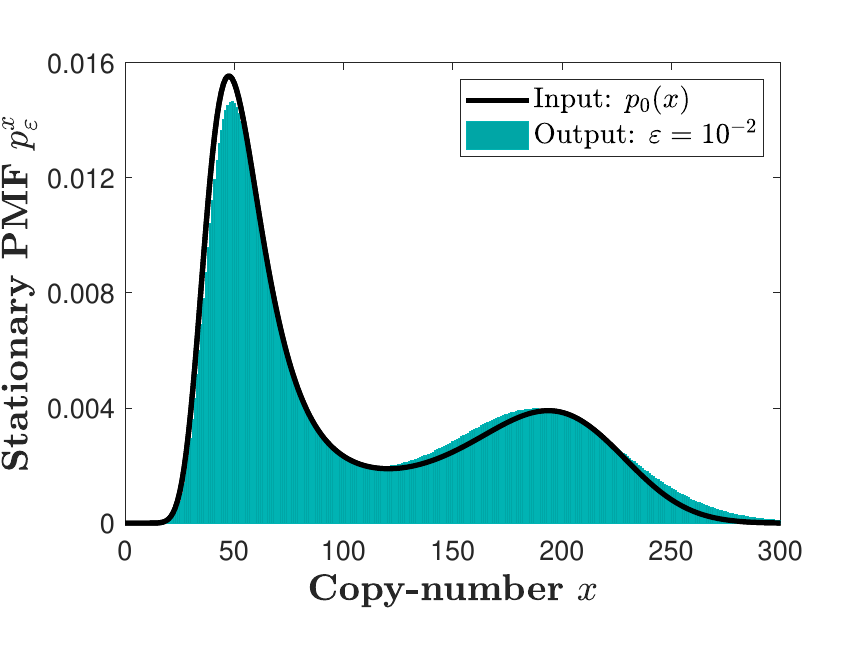}
\hskip -0.3cm
\includegraphics[width=0.4\columnwidth]{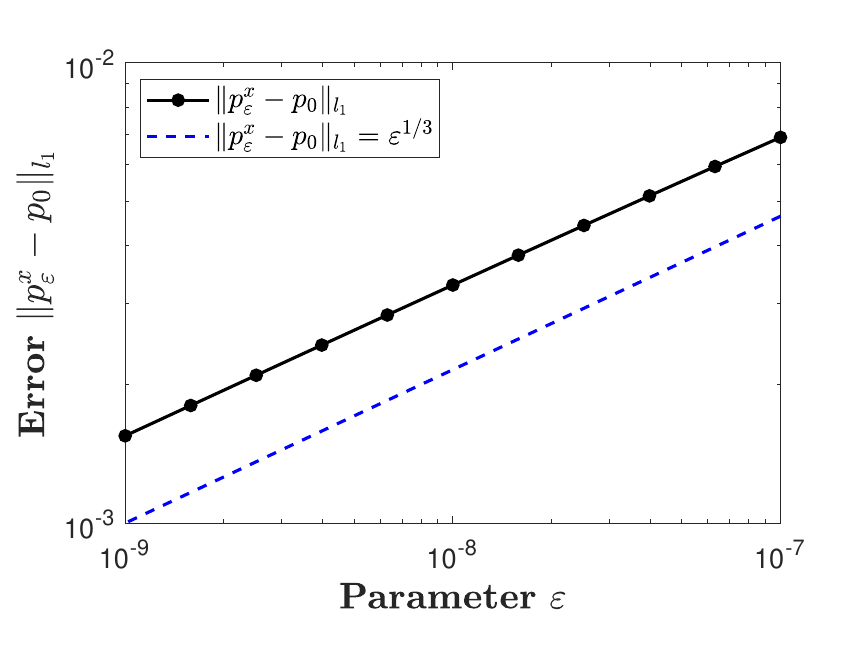}
\hskip -0.3cm
\includegraphics[width=0.4\columnwidth]{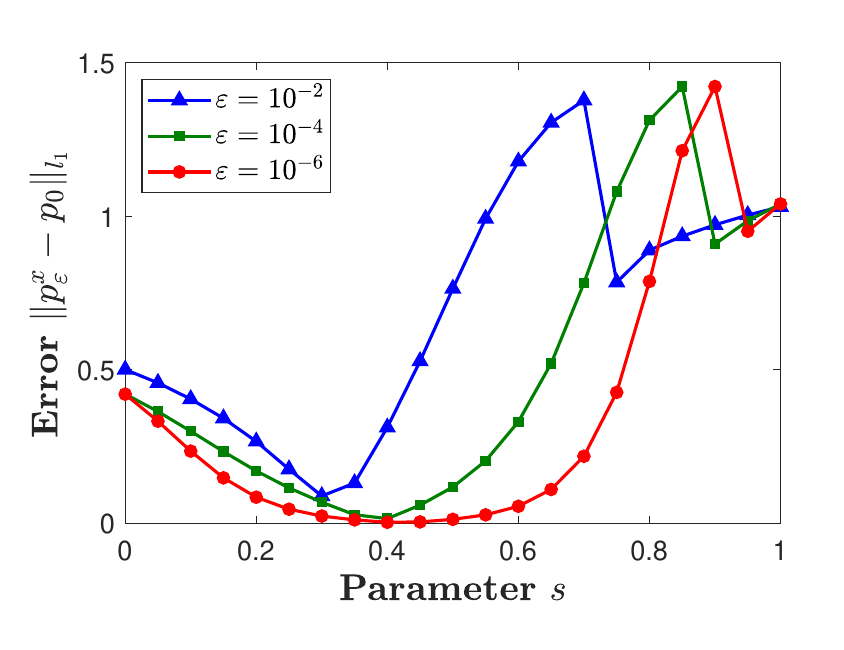}
}
\vskip -5.4cm
\leftline{\hskip -1.0cm (a) \hskip 5.8cm (b) \hskip 5.8cm (c)}
\vskip 4.4cm
\caption{\it{Panel {\rm (a)} 
displays the stationary {\rm PMF} of the input network~{\rm (\ref{eq:netschlogl})}
as a black curve, and the $x$-marginal {\rm PMF} for the output network~
{\rm (\ref{eq:netschloglmodified})}, under~{\rm(\ref{eq:scaling3_ex2})} 
with $\varepsilon = 10^{-2}$, as a blue histogram.
Panel {\rm (b)} displays a log-log plot of the $l^1$-distance between the {\rm PMF}s 
for networks~{\rm (\ref{eq:netschlogl})} and~{\rm (\ref{eq:netschloglmodified})}
as a function of $\varepsilon$. Panel {\rm (c)} displays the error 
as a function of the scaling factor $s$, 
with fixed $\bar{\kappa}_1, \bar{\kappa}_2$ and different values of $\varepsilon$.
The parameters are fixed
as in {\rm Figure~\ref{fig:Schlogl1}}.}} \label{fig:Schlogl2}
\end{figure}

Let us first satisfy kinetic condition~(\ref{eq:kinetic3}) by setting 
\begin{align}
s = 1/2, \;\; \bar{\kappa}_1 = \bar{\kappa}_2 = k_4^{1/2}.  \label{eq:scaling3_ex}
\end{align}
In Figure~\ref{fig:Schlogl1}(a), we display the stationary $x$-marginal 
PMF of the output network~(\ref{eq:netschloglmodified})
 under~(\ref{eq:scaling3_ex}), denoted by 
$p_{\varepsilon}^x = p_{\varepsilon}^x(x)$,
for different values of the parameter $\varepsilon$. 
In particular, when $\varepsilon = 10^{-2}$, the PMF
is shown as a dashed purple curve; while bi-modal, this intermediate
PMF is inaccurately distributed. On the other hand, 
when $\varepsilon = 10^{-6}$, the PMF $p_{\varepsilon}^x$
is shown as a blue histogram, and is in an excellent match with target PMF $p_0$. 
In Figure~\ref{fig:Schlogl1}(b), we show a log-log plot of a numerically
approximated error $\|p_{\varepsilon}^x - p_0\|_{l_1}$
as a function of $\varepsilon$. Also shown, as a dashed blue line, 
is the reference curve $\|p_{\varepsilon}^x - p_0 \|_1 = \varepsilon^{1/2}$;
one can notice an excellent match in the slopes of the two curves,
in accordance with the finite-time result from Proposition~\ref{proposition:convergence3}.
In Figure~\ref{fig:Schlogl2}(c), we display the $y$-marginal PMF
for network~(\ref{eq:netschloglmodified}) when $\varepsilon = 10^{-6}$, 
which is shown in Figure~\ref{fig:Schlogl2}(d) to converge, 
at $\varepsilon^{1/2}$-rate, to the Kronecker-delta distribution 
centered at zero.

Proposition~\ref{proposition:convergence3} provides information
about the error $\|p_{\varepsilon}^x - p_0\|_{l_1}$ in the limit
$\varepsilon \to 0$. Let us now discuss how one may decrease
the error for a fixed $\varepsilon$ by choosing an appropriate scaling factor $s$. 
To this end, note that network~(\ref{eq:subnetworks3}) 
consists of an ordered chain of reactions: in order for $\mathcal{R}_2$ to fire, and 
mimic~(\ref{eq:reduced3}), one requires that $\mathcal{R}_1$ fires first.
The reactant of $\mathcal{R}_1$, forming the start of the chain, is given by $2 X$, 
and the propensity function is given by $\alpha_1(x) = \kappa_1 x (x - 1)$.
On the other hand, $\mathcal{R}_2$ involves as a reactant
the short-lived lower-copy-number species $Y$, with the 
propensity function $\alpha_2(x,y_1) = \kappa_2 x y$.
Since $Y(t)$ spends most of the time at $y = 0$ for smaller $\varepsilon$, 
it follows that the underlying joint-PMF 
is concentrated in the neighborhood of the $x$-axis, and 
$\alpha_2(x,y)/\kappa_2 < \alpha_1(x)/\kappa_1$. This observation
suggests that, for a fixed $\varepsilon$, there is an optimal ratio $\kappa_1/\kappa_2$, 
sufficiently small to speed up reaction $\mathcal{R}_2$, 
and sufficiently large to ensure that reaction $\mathcal{R}_1$ 
is triggered often enough. To this end, let us now satisfy~(\ref{eq:kinetic3}) by setting 
\begin{align}
s = 1/3, \;\; \bar{\kappa}_1 = \bar{\kappa}_2 = k_4^{1/2}.  \label{eq:scaling3_ex2}
\end{align}
In Figure~\ref{fig:Schlogl2}(a), we display the stationary $x$-marginal 
PMF of the output network~(\ref{eq:netschloglmodified})
 under~(\ref{eq:scaling3_ex2}) when $\varepsilon = 10^{-2}$. 
Comparing Figures~\ref{fig:Schlogl1}(a) and~\ref{fig:Schlogl2}(a),
one can notice that, when $\varepsilon = 10^{-2}$, scaling~(\ref{eq:scaling3_ex2})
leads to a significantly better approximation than~(\ref{eq:scaling3_ex}).
However, comparing Figures~\ref{fig:Schlogl1}(b) and~\ref{fig:Schlogl2}(b),
one can notice that the convergence under the scaling~(\ref{eq:scaling3_ex})
is faster than under~(\ref{eq:scaling3_ex2}), with the latter occurring 
at a rate $\varepsilon^{1/3}$. In Figure~\ref{fig:Schlogl2}(c), we display the 
$l_1$-distance between the input and output PMFs as a function
of the scaling factor $s$, for fixed $\bar{\kappa}_1, \bar{\kappa}_2$,
and for three different values of the parameter $\varepsilon$. 
One can notice that the error is minimized approximated at $s = 3/10$
when $\varepsilon = 10^{-2}$, and that, for larger values of $\varepsilon$, 
an overall better performance is achieved by taking $s < 1/2$. 
Figure~\ref{fig:Schlogl2}(c) also suggests that the error does not 
converge to zero at the degenerate points $s = 0$ and $s = 1$. 

\section{General case: Multi-species higher-order reactions} \label{sec:nmolecular}
Let us consider an arbitrary $n$th-order reaction, under mass-action kinetics, 
with $n \ge 3$, involving $N$ biochemical species $\mathcal{X} = \{X_1, X_2, \ldots, X_N\}$
and $m$ distinct reactants $\{X_1, X_2, \ldots, X_m\} \subseteq \mathcal{X}$, given by
\begin{align}
\mathcal{R}_0(\mathcal{X}):  \; \;  \sum_{i = 1}^{m} \nu_{i} X_i  & \xrightarrow[]{k} 
\sum_{i = 1}^N \bar{\nu}_i X_i, 
\hspace{0.3cm} \textrm{where }
\{\nu_i \in \mathbb{Z}_{>}\}_{i = 1}^m
\textrm{ and }
 \sum_{i = 1}^{m} \nu_{i}  = n \ge 3.
\label{eq:reducedn}
\end{align}
For convenience, we assume that the reactant species are ordered 
according to nondecreasing stoichiometric coefficients,
$\nu_i \le \nu_j$ if $i < j$, for all $i, j \in \{1, 2,  \ldots, m\}$.
Let us also consider the second-order mass-action reaction network 
$\mathcal{R}_{\varepsilon}$, given by
\begin{align}
\mathcal{R}_{\varepsilon}(\mathcal{X}, \mathcal{Y}) & = \begin{cases}
\mathcal{R}_{1}^{\varepsilon}(X_1,X_1) \, \bigcup_{i = 2}^{\nu_1 - 2}
 \mathcal{R}_{i}^{\varepsilon}(X_1) \, \bigcup \, \mathcal{R}_{n-1}(X_1), & \text{if } m = 1, \\
\mathcal{R}_{1}^{\varepsilon}(X_1, X_2) \, \bigcup_{i = 2}^{\nu_1} 
\mathcal{R}_{i}^{\varepsilon}(X_1) \, \bigcup_{l = 2}^{m-1} 
\bigcup_{i = \delta_{l,2} + \sum_{j=1}^{l-1} \nu_j}^{-1 + \sum_{j=1}^{l} \nu_j} 
\mathcal{R}_{i}^{\varepsilon}(X_l) \\
 \,  \bigcup_{i = \delta_{m,2} + \sum_{j=1}^{m-1} \nu_j}^{n - 2} 
\mathcal{R}_{i}^{\varepsilon}(X_{m}) \, \bigcup
\,\mathcal{R}_{n-1}(X_m), & \text{if } m \ge 2,  \label{eq:fulln}
\end{cases} 
\end{align}
with the convention that $\bigcup_{l = a}^{b} \mathcal{R}(l) = \emptyset$ if $a < b$, 
where $\emptyset$ is the empty set, and with the sub-networks 
\begin{align}
\mathcal{R}_{1}^{\varepsilon}(X_i, X_j): \; \; 
X_i + X_j  & \xrightleftharpoons[1/\varepsilon]{\kappa_1}  Y_1, \nonumber \\
\mathcal{R}_{i}^{\varepsilon}(X_j): \; \; 
X_j + Y_{i-1} &\xrightleftharpoons[1/\varepsilon]{\kappa_{i}} Y_i, 
\hspace{0.3cm} \textrm{for all } i \in \{2, 3, \ldots, n-2\}, \nonumber \\
\mathcal{R}_{n-1}(X_j): \; \; 
 X_j + Y_{n-2} &\xrightarrow[]{\kappa_{n-1}} 
\sum_{i = 1}^m \tilde{\nu}_i X_i + 
\sum_{i = m+1}^N \bar{\nu}_i X_i +  
\sum_{i = 1}^{n-2} \bar{\gamma}_{i} Y_{i}. \label{eq:subnetworksn}
\end{align}
Network $\mathcal{R}_{\varepsilon} = \mathcal{R}_{\varepsilon}(\mathcal{X}, \mathcal{Y})$, 
given by~(\ref{eq:fulln})--(\ref{eq:subnetworksn}), contains 
$(n-2)$ auxiliary species $\mathcal{Y} = \{Y_1, Y_2, \ldots, Y_{n-2} \}$, 
and consists of $(2 n - 3)$ reactions, $(n-2)$ of which are first-order, 
and $(n-1)$ of second-order. Reaction network~(\ref{eq:subnetworks3}) 
from Section~\ref{sec:trimolecular} is a special case 
of~(\ref{eq:fulln})--(\ref{eq:subnetworksn}) with $n = 3$ and $m = 1$. 

\subsection{Kinetic and stoichiometric conditions} \label{sec:conditionsn}
In Appendix~\ref{app:formal}, we generalize the formal perturbation 
analysis performed for network~(\ref{eq:subnetworks3}) 
in Section~\ref{sec:perturbation3} to the 
network~(\ref{eq:fulln})--(\ref{eq:subnetworksn}), and 
show that the CMEs for input network~(\ref{eq:reducedn}) 
and output network~(\ref{eq:fulln})--(\ref{eq:subnetworksn}) match under
suitable generalizations of the kinetic and stoichiometric 
conditions~(\ref{eq:kinetic3})--(\ref{eq:stoichiometric3}). 
In particular, the generalized kinetic condition is given by
\begin{align}
\varepsilon^{n-2} \prod_{i = 1}^{n-1} \kappa_i  & =  k, 
 \; \; \; \textrm{where }
\kappa_1, \kappa_2, \ldots, \kappa_{n-1}  = o(\varepsilon^{-1}) \; \; \text{as } \varepsilon \to 0.
\label{eq:kineticn}
\end{align}
Requirement~(\ref{eq:kineticn}) states that 
the product of the rate coefficients of the slower reactions 
from~(\ref{eq:subnetworksn}), $\prod_{i = 1}^{n-1} \kappa_i$, 
divided by the product of the rate coefficient of the faster reaction, 
$1/\varepsilon^{n-2}$, must be equal to the rate coefficient of~(\ref{eq:reducedn}), $k$. 
On the other hand, when the reaction $\mathcal{R}_{n-1}(X_m)$
from~(\ref{eq:subnetworksn}) does not contain the auxiliary species
 $\{Y_1, Y_2, \ldots, Y_{n-3}\}$, then 
the stoichiometric conditions are given by
\begin{align}
\tilde{\nu}_i & =  \bar{\nu}_i - (\nu_i - \delta_{i,m}) \bar{\gamma}_{n-2}, 
\; \; \textrm{for all } i \in \{1, 2, \ldots, m \}, \; \; \; \textrm{if } 
(\bar{\gamma}_1, \bar{\gamma}_2, \ldots, \bar{\gamma}_{n-3}) = 
(0, 0, \ldots, 0). \label{eq:stoichiometricn}
\end{align}
Stoichiometric conditions valid when $(\bar{\gamma}_1, \bar{\gamma}_2, \ldots, \bar{\gamma}_{n-3})
\ne \mathbf{0}$ take a more complicated form, and can be obtained
algebraically as explained in Appendix~\ref{app:formal}.
For any particular reaction network, stoichiometric conditions are readily 
obtainable, as we now outline via an example. 

\begin{example} \label{ex:multiplecoefficients} 
Consider the fourth-order input reaction
\begin{align}
\mathcal{R}_0(X_1, X_2):  \; \;  2 X_1 + 2 X_2 & \xrightarrow[]{ k} 4 X_1 + 4 X_2,
\label{eq:ex1input}
\end{align}
with the reactant and product stoichiometric vectors $(\nu_1, \nu_2) = (2, 2)$
and $(\bar{\nu}_1, \bar{\nu}_2) = (4, 4)$, respectively, 
and the reaction vector $(\Delta  x_1, \Delta  x_2) = (4,4) - (2,2) = (2,2)$.
Output network~{\rm(\ref{eq:fulln})}--{\rm(\ref{eq:subnetworksn})} takes the
form $\mathcal{R}_{\varepsilon} = 
\mathcal{R}_{1}^{\varepsilon}(X_1,X_2) \cup \mathcal{R}_{2}^{\varepsilon}(X_1) 
\cup \mathcal{R}_{3}(X_2)$, with
\begin{align}
\mathcal{R}_{1}^{\varepsilon}(X_1,X_2): \; \; 
X_1 + X_2  & \xrightleftharpoons[1/\varepsilon]{\kappa_1}  Y_1, \nonumber \\
\mathcal{R}_{2}^{\varepsilon}(X_2): \; \; 
X_1 + Y_1  & \xrightleftharpoons[1/\varepsilon]{\kappa_2} Y_2, \nonumber \\
\mathcal{R}_{3}(X_2): \; \; 
X_2 + Y_2 & \xrightarrow[]{\kappa_3}  \tilde{\nu}_1 X_1 + \tilde{\nu}_2 X_2
+  \bar{\gamma}_{1} Y_{1} + \bar{\gamma}_{2} Y_{2}.
\label{eq:ex1output}
\end{align}
General stoichiometric conditions required for matching
networks~{\rm (\ref{eq:ex1input})}--{\rm (\ref{eq:ex1output})}
are obtained from the conservation laws that are
locally valid for the fastest two reactions from~{\rm (\ref{eq:ex1output})}:
\begin{align}
\bar{x}_1 & = x_1 + y_1 + 2 y_2, \nonumber \\
\bar{x}_2 & = x_2 + y_1 +  y_2. 
\label{eq:coordinate_transform_ex}
\end{align}
Applying the difference operator 
$\Delta$ on~{\rm (\ref{eq:coordinate_transform_ex})},
 and setting $(\Delta  \bar{x}_1, \Delta  \bar{x}_2)= (\Delta  x_1, \Delta  x_2)$, one obtains the stoichiometric conditions
\begin{align}
\tilde{\nu}_1 & = 4 - (\bar{\gamma}_1 + 2 \bar{\gamma}_2), \nonumber \\
\tilde{\nu}_2 & = 4 - (\bar{\gamma}_1 + \bar{\gamma}_2).
\label{eq:generalized_stoichiometry}
\end{align}

Conditions~{\rm(\ref{eq:generalized_stoichiometry})} can also
be applied graphically using the formal equalities 
$Y_1 \doteq (X_1 + X_2)$ and 
$Y_2 \doteq (X_1 + Y_1) = (2 X_1 + X_2)$. 
In particular, taking $(\bar{\gamma}_{1}, \bar{\gamma}_{2}) = (0, 0)$,
equation~{\rm(\ref{eq:stoichiometricn})} (or~{\rm(\ref{eq:generalized_stoichiometry})}) implies
that $(\tilde{\nu}_1, \tilde{\nu}_2) = (\bar{\nu}_1, \bar{\nu}_2)$, 
i.e. reactions $\mathcal{R}_{3}(X_2)$ from~{\rm (\ref{eq:ex1output})}
and~{\rm(\ref{eq:ex1input})} have identical products:
\begin{align}
\mathcal{R}_{3}(X_2): \; \; 
X_2 + Y_2 & \xrightarrow[]{\kappa_3}  4 X_1 + 4 X_2, 
\hspace{0.3cm} \textrm{if } (\bar{\gamma}_{1}, \bar{\gamma}_{2}) = (0, 0).
\label{eq:ex1output0}
\end{align}
One can add $\varnothing \doteq (Y_1 - X_1 - X_2)$ 
and $\varnothing \doteq (Y_2 - 2 X_1 - X_2)$ to the products 
in~{\rm (\ref{eq:ex1output0})} as many times
as desired, as long as the resulting complex contains nonnegative
stoichiometric coefficients.
For example, by adding the complex $\varnothing \doteq (Y_2 - 2 X_1 - X_2)$ once 
to~{\rm (\ref{eq:ex1output0})}, one obtains 
\begin{align}
\mathcal{R}_{3}(X_2): \; \; 
X_2 + Y_2 & \xrightarrow[]{\kappa_3}  2 X_1 + 3 X_2 + Y_2, 
\hspace{0.3cm} \textrm{if } (\bar{\gamma}_{1}, \bar{\gamma}_{2}) = (0, 1).
\label{eq:ex1output01}
\end{align}
Adding the complex 
$\varnothing \doteq (Y_1 - X_1 - X_2)$ 
four times to~{\rm (\ref{eq:ex1output0})} leads to
\begin{align}
\mathcal{R}_{3}(X_2): \; \; 
X_2 + Y_2 & \xrightarrow[]{\kappa_3}  4 Y_1,
\hspace{0.3cm} \textrm{if } (\bar{\gamma}_{1}, \bar{\gamma}_{2}) = (4, 0),
\label{eq:ex1output40}
\end{align}
while adding $\varnothing \doteq (Y_1 - X_1 - X_2)$ twice, 
and $\varnothing \doteq (Y_2 - 2 X_1 - X_2)$ once, results in
\begin{align}
\mathcal{R}_{3}(X_2): \; \; 
X_2 + Y_2 & \xrightarrow[]{\kappa_3}  X_2 + 2 Y_1 + Y_2, 
\hspace{0.3cm} \textrm{if } (\bar{\gamma}_{1}, \bar{\gamma}_{2}) = (2, 1).
\label{eq:ex1output21}
\end{align}
On the other hand, adding $\varnothing \doteq (Y_2 - 2 X_1 - X_2)$
three times to the products in~{\rm (\ref{eq:ex1output0})} leads to
\begin{align}
\mathcal{R}_{3}(X_2): \; \; 
X_2 + Y_2 & \xrightarrow[]{\kappa_3} - 2 X_1 + X_2 + 3 Y_2, 
\hspace{0.3cm} \textrm{if } (\bar{\gamma}_{1}, \bar{\gamma}_{2}) = (0, 3),
\label{eq:ex1output04}
\end{align}
which is not a chemical reaction, as the
 product complex is not nonnegative. 
\end{example}
\emph{Remark}. The approach taken in Example~\ref{ex:multiplecoefficients}
applies generally: one can extract the formal equalities, 
such as $\varnothing \doteq (Y_1 - X_1 - X_2)$ 
and $\varnothing \doteq (Y_2 - 2 X_1 - X_2)$, directly from $\mathcal{R}_{\varepsilon}$.
Writing the final reaction from $\mathcal{R}_{\varepsilon}$
with the same product complex as in the original reaction $\mathcal{R}_{0}$,
one can then add the formal equalities as many times as desired 
to the products of $\mathcal{R}_{0}$, provided the 
resulting complex remains nonnegative. 

\subsection{Convergence}
We now generalize Proposition~\ref{proposition:convergence3}, 
by establishing convergence when the slower rate coefficients
from~(\ref{eq:fulln})--(\ref{eq:subnetworksn}) are all scaled identically:
\begin{align}
\kappa_i & = \varepsilon^{-(n-2)/(n-1)} \bar{\kappa}_i, 
\; \; \; \;
\textrm{for all } i \in \{1, 2, \ldots, n-1\},
\label{eq:symmetrickinetic}
\end{align}
where $\{\bar{\kappa}_i \}_{i = 1}^{n-1}$ are $\varepsilon$-independent parameters.  
To this end, consider the PMF
\begin{align}
p_0(\mathbf{\bar{x}},\mathbf{y},t) & = p_0(\mathbf{\bar{x}},t) 
\prod_{i=1}^{n-2} \delta_{y_i,0},
 \label{eq:solnp0_scaled}
\end{align}
where $\mathbf{\bar{x}} \in \mathbb{Z}_{\ge}^N$ 
is given by~(\ref{eq:newvariablesn1})--(\ref{eq:newvariablesn}) in Appendix~\ref{app:formal},
$\mathbf{y} \in \mathbb{Z}_{\ge}^{n-2}$,
and the PMF $p_0(\mathbf{\bar{x}},t)$ satisfies 
\begin{align}
 \frac{\mathrm{d}}{\mathrm{d} t} p_0(\mathbf{\bar{x}},t)  & = 
\left(E_{\mathbf{\bar{x}}}^{-\Delta \bar{\mathbf{x}}} - 1 \right) 
\left(\prod_{j = 1}^{n-1} \bar{\kappa}_j \right)
\prod_{l = 1}^m \bar{x}_l^{\underline{\nu_l}} p_0(\mathbf{\bar{x}},t), 
\label{eq:effectiveCMEn_scaled}
\end{align}
where $\Delta \mathbf{\bar{x}}$ is obtained
by applying the difference operator 
$\Delta$ on~(\ref{eq:newvariablesn1})--(\ref{eq:newvariablesn}). 
Note that, under the kinetic and stoichiometric 
conditions from Section~\ref{sec:conditionsn},
CME~(\ref{eq:effectiveCMEn_scaled})
is identical to the CME of the original network~(\ref{eq:reducedn}).

\begin{theorem} \label{theorem:convergencen} 
\textit{Consider network $\mathcal{R}_{\varepsilon}$, 
given by~{\rm(\ref{eq:fulln})}--{\rm(\ref{eq:subnetworksn})}, 
with the rate coefficients $\{\kappa_i\}_{i = 1}^{n-1}$ given
by~{\rm (\ref{eq:symmetrickinetic})}, and with the {\rm PMF} 
$p_{\varepsilon}(\mathbf{\bar{x}},\mathbf{y},t)$ satisfying {\rm(\ref{eq:rescaledCMEn})}.
Let the {\rm PMF} $p_0(\mathbf{\bar{x}},\mathbf{y},t)$ 
satisfy~{\rm (\ref{eq:solnp0_scaled})}--{\rm (\ref{eq:effectiveCMEn_scaled})}, 
and assume that $p_{\varepsilon}(\mathbf{\bar{x}},\mathbf{y},0) = p_0(\mathbf{\bar{x}},\mathbf{y},0)$.
Then, for every $\mathbb{S} \subset \mathbb{Z}_{\ge}^{N + (n-2)}$
and every $T > 0$ there exists an $\varepsilon$-independent constant $c > 0$ such that
\begin{align}
 \left \|p_{\varepsilon}(\mathbf{\bar{x}},\mathbf{y},t) 
- p_0(\mathbf{\bar{x}},\mathbf{y},t) \right \|_{l_1(\mathbb{S})} & 
\le  c \, \varepsilon^{1/(n-1)}, \; \;  \textrm{as } \varepsilon \to 0,
\; \; \textrm{for all } t \in [0,T].
\label{eq:convergencen}
\end{align}
}
\end{theorem} 

\begin{proof}
See Appendix~\ref{app:convergence}.
\end{proof}
\noindent \emph{Remark}. The error estimate~(\ref{eq:convergencen}) 
is independent of how the stoichiometric coefficients $\{\tilde{\nu}_i\}_{i = 1}^{m}$, 
$\{\bar{\nu}_i\}_{i = m+1}^{N}$ and $\{\bar{\gamma}_i\}_{i = 1}^{n-2}$
from~(\ref{eq:subnetworksn}) are chosen. 

To formulate Theorem~\ref{theorem:convergencen}, 
we have assumed a fixed ordering of the reactants and reactions 
in~(\ref{eq:fulln})--(\ref{eq:subnetworksn}). One can readily
prove analogous results for other suitable orderings.

\begin{example} \label{ex:multipleoutput} 
Consider the third-order input reaction
\begin{align}
\mathcal{R}_0(X_1, X_2):  \; \;  X_1 + 2 X_2 & \xrightarrow[]{k}  \varnothing. 
\label{eq:ex2input}
\end{align}
Output network~{\rm(\ref{eq:fulln})}--{\rm(\ref{eq:subnetworksn})} is given 
by  $\mathcal{R}_{\varepsilon} = 
\mathcal{R}_{1}^{\varepsilon}(X_1,X_2) \cup \mathcal{R}_{2}(X_2)$, where
\begin{align}
\mathcal{R}_{1}^{\varepsilon}(X_1,X_2): \; \; 
X_1 + X_2  \xrightleftharpoons[1/\varepsilon]{\kappa_1}  Y_1, \nonumber \\
\mathcal{R}_{2}(X_2): \; \; 
X_2 + Y_1 \xrightarrow[]{\kappa_2}  \varnothing;
\label{eq:ex2output1}
\end{align}
in particular, the forward reaction from $\mathcal{R}_{1}^{\varepsilon}$
is a second-order hetero-reaction, involving two distinct reactants $X_1$ and $X_2$. 
One can readily show that the results presented in this 
section also hold for the output network
$\mathcal{R}_{1}^{\varepsilon}(X_2,X_2) \cup \mathcal{R}_{2}(X_1)$,
given by
\begin{align}
\mathcal{R}_{1}^{\varepsilon}(X_2,X_2): \; \; 
2 X_2  \xrightleftharpoons[1/\varepsilon]{\kappa_1} Y_1, \nonumber \\
\mathcal{R}_{2}(X_1): \; \; 
X_1 + Y_1 \xrightarrow[]{\kappa_2}  \varnothing,
\label{eq:ex2output2}
\end{align}
for which the forward reaction from $\mathcal{R}_{1}^{\varepsilon}$
 is a second-order homo-reaction, involving $X_2$.
Another valid output network is $\mathcal{R}_{1}^{\varepsilon}(X_1, \varnothing) \cup 
\mathcal{R}_{2}^{\varepsilon}(X_2) 
\cup \mathcal{R}_{3}(X_2)$, given by
\begin{align}
\mathcal{R}_{1}^{\varepsilon}(X_1, \varnothing): \; \; 
X_1 \xrightleftharpoons[1/\varepsilon]{\kappa_1}  Y_1, \nonumber \\
\mathcal{R}_{2}^{\varepsilon}(X_2): \; \; 
X_2 + Y_1 \xrightleftharpoons[1/\varepsilon]{\kappa_2}  Y_2, \nonumber \\
\mathcal{R}_{3}(X_1): \; \; 
X_2 + Y_2 \xrightarrow[]{\kappa_3}  \varnothing,
\label{eq:ex2output2}
\end{align}
for which the forward reaction from $\mathcal{R}_{1}^{\varepsilon}$ is of first-order.
\end{example}

\section{Examples: Noise-induced phenomena} \label{sec:examples}
In this section, we apply the results from Section~\ref{sec:nmolecular}
to two test networks arising from theoretical synthetic biology
and displaying noise-induced dynamical phenomena.
In particular, the first network, given by~(\ref{eq:exKronecker}),
plays an important role in the stochastic morpher controller~\cite{MeRobust}
that can globally morph PMF of a given reaction network into any desired form.
The second network, given by~(\ref{eq:nettristable}), is a part
of the noise-control algorithm~\cite{Me3} that can redesign
a given reaction network to locally reshape the underlying PMF in 
a mean-preserving manner. 

\subsection{Biochemical Kronecker-delta distribution} \label{sec:Kronecker}
\begin{figure}[!htbp]
\vskip 0.2cm
\centerline{
\hskip 1mm
\includegraphics[width=0.4\columnwidth]{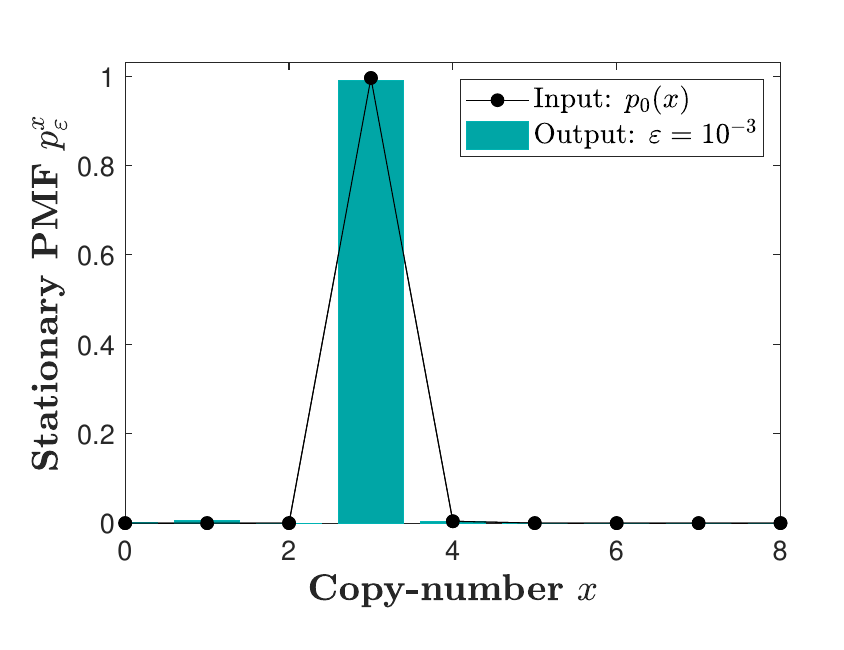}
\hskip -0.3cm
\includegraphics[width=0.4\columnwidth]{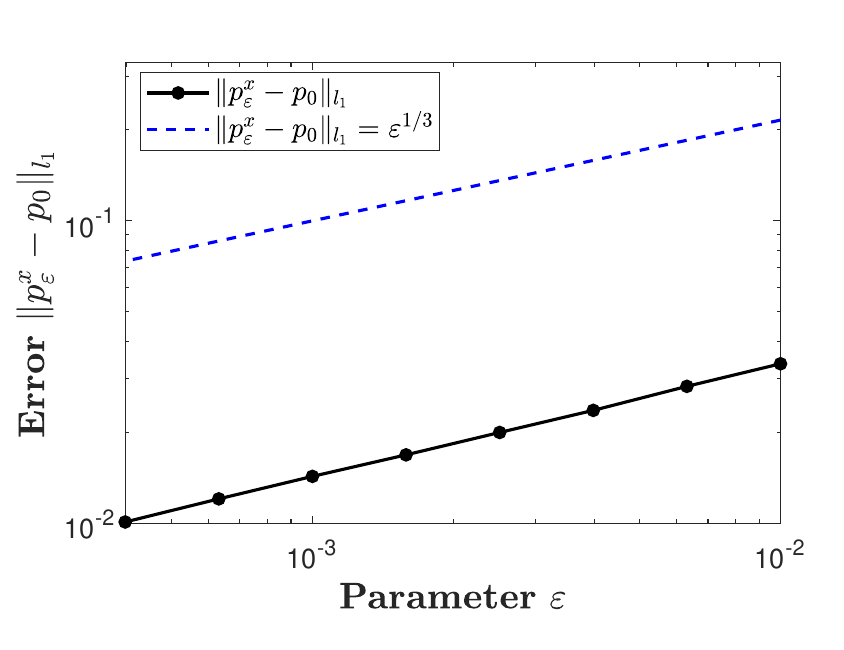}
\hskip -0.3cm
\includegraphics[width=0.4\columnwidth]{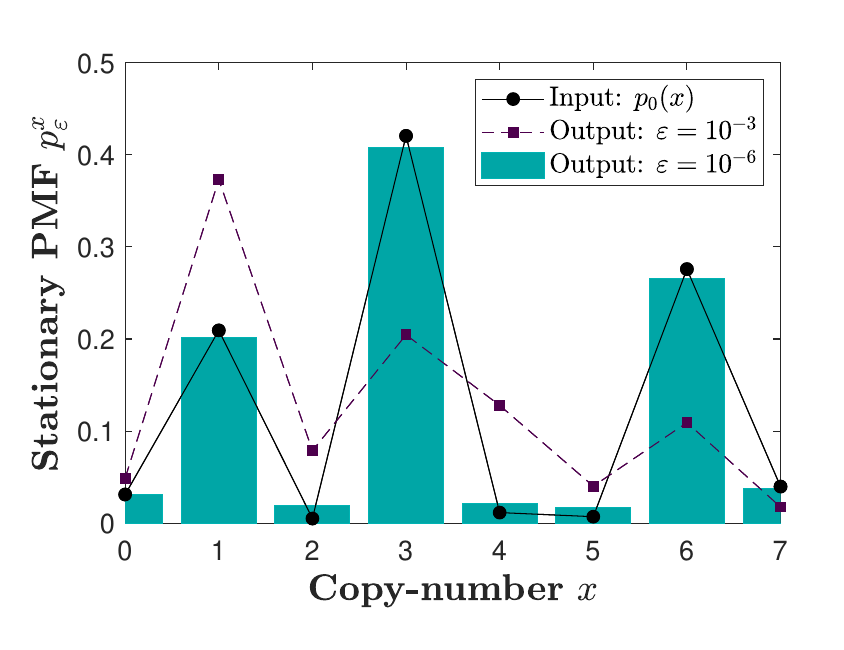}
}
\vskip -5.4cm
\leftline{\hskip -0.8cm (a) \hskip 5.7cm (b) \hskip 5.8cm (c)}
\vskip 4.5cm
\caption{
{\it  Panel {\rm (a)} displays the long-time {\rm PMF} of the 
input network~{\rm (\ref{eq:exKronecker})} with $(k_1, k_2) = (10^{-3}, 10^{-2})$
as black dots interpolated with solid lines; the long-time 
$x$-marginal {\rm PMF} of the output 
network~{\rm (\ref{eq:exKroneckerappr})}, with 
$(\tilde{\nu}, \bar{\gamma}_1,\bar{\gamma}_2) = (0,0,1)$, 
under rate coefficients~{\rm (\ref{eq:Kroneckercoeff})}, 
is shown as a blue histogram when $\varepsilon = 10^{-2}$.
Panel {\rm (b)} displays a log-log plot of the $l^1$-distance between 
the long-time {\rm PMF}s for 
networks~{\rm (\ref{eq:exKronecker})} and~{\rm (\ref{eq:exKroneckerappr})}.
Panel {\rm (c)} displays the long-time {\rm PMF} of the 
input network~{\rm (\ref{eq:nettristable})} with 
$(k_1, k_2, k_{2,5}, \tilde{k}_{4,2}) = (1,1,1,1)$,
and the $x$-marginal {\rm PMF} of the
 output network~{\rm (\ref{eq:nettristableoutput})}, 
under rate coefficients~{\rm (\ref{eq:ttristabk})} with $\beta = 1/12$,
for two different values of $\varepsilon$. 
The conservation constant for~{\rm (\ref{eq:nettristable})}
 is fixed to $c = 7$, and~{\rm (\ref{eq:nettristableoutput})}
is initialized with zero copy-numbers of the species
$\{Y_i\}_{i = 1}^5$ and $\tilde{Y}_4$.}
}
\label{fig:Noise}
\end{figure}

Let us consider the fourth-order mass-action input reaction network 
\begin{align}
\mathcal{R}_0(X): \; \;  
\varnothing & \xrightarrow[]{k_1} X, \nonumber \\
4 X & \xrightarrow[]{k_2} 3 X. \label{eq:exKronecker}
\end{align}
Long-time PMF of~(\ref{eq:exKronecker}), under a particular choice
of the rate coefficients $k_1 < k_2$, is shown in Figure~\ref{fig:Noise}(a)
as black dots interpolated with solid lines. 
The PMF is approximately the Kronecker-delta distribution centered $x = 3$.
In particular, when there are less than four molecules of $X$ 
present, $x < 4$, only the first reaction from~(\ref{eq:exKronecker}) fires
and $X$ experiences a constant positive drift until four molecules 
are present. When $x \ge 4$, both reactions from~(\ref{eq:exKronecker}) fire, 
with the second one, having a larger propensity function, 
overpowering the first one and generating a net-negative drift.
The combined effect of the two reactions forces 
$X$ to spend most of the time at the state $x = 3$. 

The fourth-order input network~(\ref{eq:exKronecker}) 
can be approximated with a second-order output one using the results from 
Section~\ref{sec:nmolecular}. In particular, applying the
 algorithm~(\ref{eq:fulln})--(\ref{eq:subnetworksn}), 
a suitable output network is given by 
$\mathcal{R} \cup \mathcal{R}_{1}^{\varepsilon} 
\cup \mathcal{R}_2^{\varepsilon} \cup \mathcal{R}_4$, where
\begin{align}
& \; \, \mathcal{R}(X):  \; 
& \varnothing & \xrightarrow[]{k_1} X, \nonumber \\
& \mathcal{R}_{1}^{\varepsilon}(X): \;  
& 2 X & \xrightleftharpoons[1/\varepsilon]{\kappa_1} Y_1, \nonumber \\
& \mathcal{R}_{2}^{\varepsilon}(X): \; 
 & X + Y_1  & \xrightleftharpoons[1/\varepsilon]{\kappa_2} Y_2, \nonumber \\
& \mathcal{R}_4(X): \;  
& X + Y_2 & \xrightarrow[]{\kappa_3} \tilde{\nu} X + \bar{\gamma}_1 Y_1 + \bar{\gamma}_2 Y_2.
\label{eq:exKroneckerappr}
\end{align}
Let us fix $(\bar{\gamma}_1,\bar{\gamma}_2) = (0,1)$, 
so that $\tilde{\nu} = 0$ by the stoichiometric condition~(\ref{eq:stoichiometricn}). 
On the other hand, the kinetic condition~(\ref{eq:kineticn})
for~(\ref{eq:exKroneckerappr}) takes the form
\begin{align}
\varepsilon^2 \kappa_1 \kappa_2 \kappa_3 & = k_{2}, 
\; \; \; \textrm{where }
\kappa_1, \kappa_2, \kappa_3, \kappa_{4}  = o(\varepsilon^{-1}) \; \; \text{as } \varepsilon \to 0,
\label{eq:kineticKronecker}
\end{align} 
which is satisfied with e.g. 
\begin{align}
\kappa_1 & =k_2 \varepsilon^{-2/3}, \; \; 
\kappa_2 = \varepsilon^{-2/3}, \; \; 
\kappa_3 =  \varepsilon^{-2/3}.
\label{eq:Kroneckercoeff}
\end{align} 
In Figure~\ref{fig:Noise}(a), we display the long-time $x$-marginal PMF 
of network~(\ref{eq:exKroneckerappr}) under the rate coefficients~(\ref{eq:Kroneckercoeff})
with $\varepsilon = 10^{-3}$, which is in an excellent agreement with the input PMF. 
In Figure~\ref{fig:Noise}(b), we show that the output PMF converges to the input 
one at a rate $\varepsilon^{1/3}$, consistent with 
Theorem~\ref{theorem:convergencen}.

\subsection{Noise-induced tri-modality} \label{sec:tristability}
In Section~\ref{sec:nmolecular}, we have provided an algorithm for
approximating a single higher-order input reaction~(\ref{eq:reducedn})
with a second-order network of the form~(\ref{eq:fulln})--(\ref{eq:subnetworksn}).
A generalization to the case when the input network 
contains multiple higher-order reactions is straightforward:
each of the  desired higher-order reaction can be separately 
approximated with a suitable network 
of the form~(\ref{eq:fulln})--(\ref{eq:subnetworksn}), 
while the other input reactions, which one does not wish 
to approximate, are copied directly to the output network 
without any modifications.
However, such an output network may be biochemically
expensive to engineer, as it may contain a larger number of the
auxiliary reactions and species $\mathcal{Y}$. 
This undesirable feature can be reduced when
some of the higher-order input reactions involve common
reactant sub-complexes; then, some of the intermediate species
$\mathcal{Y}$ may be re-used to 
simultaneously approximate multiple input reactions. 
Put it another way, reactions with common sub-complexes
can be approximated by multiple chains of reactions of the form~(\ref{eq:subnetworksn}),
which all branch out from a suitable common sub-chain, provided
the underlying kinetic conditions can be satisfied. 
These statements readily follow from the perturbation analysis
performed in this paper. To illustrate these ideas, 
consider the seventh-order mass-action input network
$\mathcal{R}_0 = \mathcal{R} \cup \mathcal{R}_{2,5}
\cup \mathcal{R}_{4,2}$, given by
\begin{align}
& \mathcal{R}(X_1, X_2): \;  & X_2  &
\xrightleftharpoons[k_2]{ k_1 } X_1, \nonumber \\
& \mathcal{R}_{2,5}(X_1, X_2): \;  & 2 X_1 + 5 X_2 &
\xrightarrow[]{ k_{2,5} } 3 X_1 + 4 X_2, \nonumber \\
 & & 2 X_1 + 5 X_2  &
\xrightarrow[]{ k_{2,5} } X_1 + 6 X_2, \nonumber \\
& \mathcal{\tilde{R}}_{4,2}(X_1, X_2): \;  & 4 X_1 + 2 X_2 &
\xrightarrow[]{\tilde{k}_{4,2} } 5 X_1 + X_2, \nonumber \\
 & & 4 X_1 + 2 X_2&
\xrightarrow[]{\tilde{k}_{4,2} } 3 X_1 + 3 X_2. 
\label{eq:nettristable}
\end{align} 
Species $X_1$ and $X_2$ from~(\ref{eq:nettristable})
are conserved, $x_1 + x_2 = c$; in what follows, 
we fix the conservation constant to $c = 7$.
Reaction network~(\ref{eq:nettristable}) has been obtained by 
applying the noise-control algorithm~\cite{Me3}
on network $\mathcal{R}(X_1,X_2)$; in particular, sub-networks $\mathcal{R}_{2,5}$
and $\mathcal{\tilde{R}}_{4,2}$, called zero-drift networks, 
introduce a state-dependent noise, and decrease the PMF
of network $\mathcal{R}(X_1,X_2)$, at $x_1 = 2$ and $x_1 \in \{4,5\}$, respectively,
while preserving the underlying mean. In Figure~\ref{fig:Noise}(c), we display the
long-time PMF of~(\ref{eq:nettristable}) as black dots interpolated with solid lines. 
One can notice that the network displays noise-induced tri-modality,
with the modes $x_1 \in \{1, 3, 6\}$. 

Using~(\ref{eq:fulln})--(\ref{eq:subnetworksn}) to independently reduce order of 
the network $\mathcal{R}_{2,5} \cup \mathcal{\tilde{R}}_{4,2}$ from seven to two
requires $18$ auxiliary species and $40$ reactions in total, i.e. $10$ independent auxiliary species 
for the network $\mathcal{R}_{2,5}$ ($5$ species for each of the underlying reactions)
and, similarly, $8$ species for $\mathcal{\tilde{R}}_{4,2}$. 
However, since each of the two reactions from $\mathcal{R}_{2,5}$ 
 involves the same reactants, one can reduce their order simultaneously
by using $5$ auxiliary species; similarly, $4$ auxiliary species
suffice to reduce order of the network $\mathcal{\tilde{R}}_{4,2}$. 
Furthermore, all of the reactions from $\mathcal{R}_{2,5} \cup \mathcal{\tilde{R}}_{4,2}$
involve a common reactant sub-complex $(2 X_1 + 2 X_2)$, so that, 
instead of using $9$ auxiliary species for $\mathcal{R}_{2,5} \cup \mathcal{\tilde{R}}_{4,2}$, 
one can use only $6$; a resulting output network is given by 
\begin{align}
& \mathcal{R}(X_1, X_2): \;  & X_2  &
\xrightleftharpoons[k_2]{ k_1 } X_1, \nonumber \\
& \mathcal{R}_{1}^{\varepsilon}(X_1, X_1): \;  & 2 X_1  &
\xrightleftharpoons[1/\varepsilon]{\kappa_1} Y_1, \nonumber \\
& \mathcal{R}_{2}^{\varepsilon}(X_2): \;  & X_2 + Y_1  &
\xrightleftharpoons[1/\varepsilon]{\kappa_2} Y_2, \nonumber \\
& \mathcal{R}_{3}^{\varepsilon}(X_2): \;  & X_2 + Y_2  &
\xrightleftharpoons[1/\varepsilon]{\kappa_3} Y_3, \nonumber \\
& \mathcal{R}_{4}^{\varepsilon}(X_2): \;  & X_2 + Y_3  &
\xrightleftharpoons[1/\varepsilon]{\kappa_4} Y_4, \nonumber \\
& \mathcal{R}_{5}^{\varepsilon}(X_2): \;  & X_2 + Y_4  &
\xrightleftharpoons[1/\varepsilon]{\kappa_5} Y_5, \nonumber \\
& \mathcal{R}_6(X_1): \;  & X_2 + Y_5 &
\xrightarrow[]{\kappa_6} 3 X_1 + 4  X_2, \nonumber \\
 & & X_2 + Y_5  &
\xrightarrow[]{\kappa_6} X_1 + 6 X_2, \nonumber \\
& \tilde{\mathcal{R}}_{4}^{\varepsilon}(X_1): \;  & X_1 + Y_3  &
\xrightleftharpoons[1/\varepsilon]{\tilde{\kappa}_4} \tilde{Y}_4, \nonumber \\
& \tilde{\mathcal{R}}_5(X_1): \;  & X_1 + \tilde{Y}_4 &
\xrightarrow[]{\tilde{\kappa}_5} 5 X_1 + X_2, \nonumber \\
 & & X_1 +\tilde{Y}_4  &
\xrightarrow[]{\tilde{\kappa}_5} 3 X_1 + 3 X_2. 
\label{eq:nettristableoutput}
\end{align} 
The sub-network $\mathcal{R}_{2,5}$ is approximated by 
$\bigcup_{i=1}^5 \mathcal{R}_{i}^{\varepsilon} \cup \mathcal{R}_6$,
while $\mathcal{\tilde{R}}_{4,2}$ by $\bigcup_{i=1}^3 \mathcal{R}_{i}^{\varepsilon} 
\cup \tilde{\mathcal{R}}_{4}^{\varepsilon} \cup \tilde{\mathcal{R}}_5$.
Instead of using $18$ auxiliary species and $40$ reactions to
independently approximate all the higher-order reactions from~(\ref{eq:nettristable}) 
by second-orders ones, we have achieved the same goal 
in~(\ref{eq:nettristableoutput}) with $6$ auxiliary species and $16$ reactions.

Network~(\ref{eq:nettristableoutput}) satisfies the 
stoichiometric conditions~(\ref{eq:stoichiometricn}), 
since $\mathcal{R}_6$ and $\tilde{\mathcal{R}}_5$ do not
contain any auxiliary species $\mathcal{Y} = \{Y_1, Y_2, Y_3, Y_4, Y_5, \tilde{Y}_4\}$
as products. One can reduce the product stoichiometric coefficients of $X_1$ and $X_2$ in 
$\mathcal{R}_6$ and $\tilde{\mathcal{R}}_5$ by introducing suitable 
species $\mathcal{Y}$ as products (see also Example~\ref{ex:multiplecoefficients}); 
for simplicity, we consider the form~(\ref{eq:nettristableoutput}) in this paper. On the other
hand, kinetic conditions~(\ref{eq:kineticn}) take the form
\begin{align}
\varepsilon^5 \kappa_1 \kappa_2 \kappa_3 
\kappa_4 \kappa_5 \kappa_6  & = k_{2,5}, 
\; \; \; \textrm{where }
\kappa_1, \kappa_2, \ldots, \kappa_6 = o(\varepsilon^{-1}) \; \; \text{as } \varepsilon \to 0,
\nonumber \\
\varepsilon^4 \kappa_1 \kappa_2 \kappa_3 
\tilde{\kappa}_4 \tilde{\kappa}_5  & = \tilde{k}_{4,2}, 
\; \; \; \textrm{where } \tilde{\kappa}_4, \tilde{\kappa}_5
= o(\varepsilon^{-1}) \; \; \text{as } \varepsilon \to 0.
\label{eq:kineticTrimodal}
\end{align} 
Guided by the discussion in Section~\ref{sec:Schlogl} (see also 
Figure~\ref{fig:Schlogl2}(c)), to achieve a higher accuracy for larger
values of $\varepsilon$, we satisfy the kinetic conditions with
\begin{align}
\kappa_1 & = \left(\varepsilon^{-\frac{5}{6}} (k_{2,5})^{\frac{1}{6}} \right)
\varepsilon^{5 \beta}, \nonumber \\
\kappa_i & = \left( \varepsilon^{-\frac{5}{6}} (k_{2,5})^{\frac{1}{6}} \right)
\varepsilon^{-\beta}, \; \; 0 \le \beta < 1/6, \; \; \textrm{for all }  i \in \{2, 3, 4, 5, 6\}, \nonumber \\
\tilde{\kappa}_i & = \varepsilon^{-2} (\kappa_1 \kappa_2 \kappa_3)^{-\frac{1}{2}} 
(\tilde{k}_{4,2})^{\frac{1}{2}}, \; \; \textrm{for all } i \in \{4, 5\}.
\label{eq:ttristabk}
\end{align} 
In Figure~\ref{fig:Noise}(c), we display
 the long-time $x$-marginal PMF of~(\ref{eq:nettristableoutput})
with rate coefficients~(\ref{eq:ttristabk}), with the auxiliary parameter $\beta = 1/12$,
for $\varepsilon = 10^{-3}$ and $\varepsilon = 10^{-6}$, the latter of which 
is in a good agreement with the long-time PMF of the input network~(\ref{eq:nettristable}).

\section{Discussion} \label{sec:discussion}
In this paper, we have shown that, by introducing suitable time-scale separations
and dimension expansion (introduction of auxiliary reactions and species), 
any higher-order input reaction can be mapped
to a suitable second-order output network, with the underlying 
stochastic dynamics being preserved.
In particular, the time-dependent probability distributions of the input 
and output networks are arbitrarily close over bounded time-intervals in an 
appropriate asymptotic limit of some of the underlying rate coefficients. 
The results presented in this paper generalize 
the deterministic order-reduction algorithm for 
third- and fourth-order reactions from~\cite{Highmol1,Highmol2,UNI3} 
to the stochastic regime and to reactions with arbitrary order.

In Section~\ref{sec:trimolecular}, we have shown that an arbitrary one-species
 reaction of order $n = 3$, given by~(\ref{eq:reduced3}), 
can be approximated by a family of second-order networks~(\ref{eq:subnetworks3}),
provided the kinetic and stoichiometric 
conditions~(\ref{eq:kinetic3}) and~(\ref{eq:stoichiometric3}), respectively, are satisfied. 
Convergence for a family of the output networks
has been proved in Proposition~\ref{proposition:convergence3}. 
In Section~\ref{sec:Schlogl}, we have verified the results from 
Section~\ref{sec:trimolecular} on the Schl\"{o}gl network~(\ref{eq:netschlogl}), 
and we have discussed different ways to satisfy the kinetic condition~(\ref{eq:kinetic3}).
In Appendices~\ref{app:formal}--\ref{app:convergence}, we have generalized the results from
Section~\ref{sec:trimolecular} to arbitrary multi-species reactions of order $n \ge 3$, 
and have presented these results in Section~\ref{sec:nmolecular}.
In particular, we have shown that any reaction of the form~(\ref{eq:reducedn})
can be approximated with a family of second-order 
networks~(\ref{eq:fulln})--(\ref{eq:subnetworksn}) that satisfies
the generalized kinetic and stoichiometric 
conditions~(\ref{eq:kineticn})--(\ref{eq:stoichiometricn}); convergence
for a family of such networks is presented as Theorem~\ref{theorem:convergencen}. 
For an input network of order $n \ge 3$, the order of convergence is shown 
to be given by $1/(n-1)$, and is independent of the stoichiometry of the output networks. 
In Section~\ref{sec:examples}, we have applied the results
from Section~\ref{sec:nmolecular} to the fourth- and seventh-order input 
networks~(\ref{eq:exKronecker}) and~(\ref{eq:nettristable}), respectively,
arising from theoretical synthetic biology~\cite{Me3,MeRobust}, and displaying
noise-induced phenomena that are absent at the deterministic level. 
In this context, we have discussed how multiple higher-order input reactions 
can be mapped more efficiently into smaller output networks.

The results established in this paper may play an important role in
synthetic biology, and particularly in nucleic-acid-based synthetic biology, 
also known as DNA computing~\cite{Experiment5}. In this context, it has been proved
that, assuming one can experimentally vary reaction rate coefficients over a sufficiently large range, 
any abstract second-order reaction network, under mass-action kinetics, can be experimentally
compiled into a physical second-order network with DNA molecules, with the
underlying deterministic dynamics being preserved over bounded time-intervals~\cite{DNAComputing1}. 
This molecular compiler has been proved to also preserve the underlying 
stochastic dynamics~\cite{Me3}. In this context, results from Section~\ref{sec:nmolecular}, 
and Theorem~\ref{theorem:convergencen} in particular, imply the following corollary.

\begin{corollary}{\rm (Universal molecular compiler)} \label{corollary:DNAcomputing} 
\textit{Assume that the reaction rate coefficients in the {\rm DNA} compiler
from~{\rm \cite{DNAComputing1}} can be varied over arbitrarily large range.
Then, mass-action input reaction networks of any order can be 
compiled into second-order {\rm DNA}-based output networks
via the compiler from~{\rm \cite{DNAComputing1}}, 
in such a way that the probability distributions for the input and output networks
are arbitrarily close over any bounded time-interval.}
\end{corollary}

\appendix

\section{Appendix: Background} \label{app:background}
\emph{Notation}. 
Union and intersection of sets $\mathcal{A}_1$ and $\mathcal{A}_2$
are denoted by $\mathcal{A}_1 \cup \mathcal{A}_2$ and 
$\mathcal{A}_1 \cap \mathcal{A}_2$, respectively.
The empty set is denoted by $\emptyset$. 
Set $\mathbb{R}$ is the space of real numbers, 
$\mathbb{R}_{\ge}$ the space of nonnegative real numbers,
 and $\mathbb{R}_{>}$ the space of positive real numbers. 
Similarly, $\mathbb{Z}$ is the space of integer numbers, 
$\mathbb{Z}_{\ge}$ the space of nonnegative integer numbers, 
and $\mathbb{Z}_{>}$ the space of positive integer numbers. 
Euclidean row-vectors are denoted in boldface, 
$\mathbf{x} = (x_1, x_2, \ldots, x_N) \in \mathbb{R}^{N} = \mathbb{R}^{1 \times N}$;
the zero vector is given by $\mathbf{0} = (0, 0, \ldots, 0) \in \mathbb{R}^N$.
Given two sequences  $p, q : \mathbb{Z}_{\ge}^N \to \mathbb{R}$, 
their $l^2$ inner-product is given by 
$\langle p(\mathbf{x}), q(\mathbf{x}) \rangle_{\mathbf{x}} =
\sum_{\mathbf{x} \in \mathbb{Z}_{\ge}^N} p(\mathbf{x}) q(\mathbf{x})$;
the $l^1$-norm of $p(\mathbf{x})$ is given by 
$ \| p(\mathbf{x}) \|_{l_1} 
= \sum_{\mathbf{x} \in \mathbb{Z}_{\ge}^N} |p(\mathbf{x})|$. 
Function $\delta_{\cdot,x_0} : \mathbb{Z} \to [0,1]$ 
with parameter $x_0 \in \mathbb{Z}$, 
defined by $\delta_{x,x_0} = 1$ if $x = x_0$, and 
$\delta_{x,x_0} = 0$ if $x \ne x_0$, is called the Kronecker-delta 
distribution centered at $x_0$.
Given two functions $f, g : \mathbb{R}_{>} \to \mathbb{R}$, 
we write $f(\varepsilon) = o(g(\varepsilon))$ as $\varepsilon \to 0$
if $\lim_{\varepsilon \to 0} f(\varepsilon)/g(\varepsilon) = 0$;
we write $f(\varepsilon) = O(g(\varepsilon))$ as $\varepsilon \to 0$
if $\lim_{\varepsilon \to 0} |f(\varepsilon)|/|g(\varepsilon)| < \infty$.

\subsection{Biochemical reaction networks} \label{app:CRNs}
We consider reaction networks $\mathcal{R} = \mathcal{R}(\mathcal{X})$
firing in well-mixed unit-volume reactors under mass-action kinetics~\cite{Feinberg},
 involving $N$ biochemical species $\mathcal{X} = \{X_1, X_2, \ldots, X_N\}$ interacting via $M$ reactions given by
\begin{align}
\mathcal{R}(\mathcal{X}): \; \; \sum_{i = 1}^{N} \nu_{j,i} X_i &\xrightarrow[]{ k_j}  
\sum_{i = 1}^N \bar{\nu}_{j,i} X_i, \, \, \, \, \, \, \,  j \in \{1, 2, \ldots, M\}. \label{eq:reactionnetworks}
\end{align}
Here, $k_j \in \mathbb{R}_{>}$  is the \emph{rate coefficient} of the $j$-reaction, 
and we let $\mathbf{k} \equiv (k_1, k_2, \ldots, k_M) \in \mathbb{R}_{>}^M$.
Integers $\nu_{j, l}, \bar{\nu}_{j, l} \in \mathbb{Z}_{\ge}$ 
are the \emph{reactant} and \emph{product stoichiometric coefficients}
of the species $X_l$ in the $j$-reaction, respectively, and we let
$\boldsymbol{\nu}_j \equiv (\nu_{j,1}, \nu_{j,2}, \ldots, \nu_{j,N}) \in \mathbb{Z}_{\ge}^N$
and $\bar{\boldsymbol{\nu}}_j \equiv (\bar{\nu}_{j,1}, \bar{\nu}_{j,2}, \ldots, \bar{\nu}_{j,N}) 
\in \mathbb{Z}_{\ge}^N$. 
If $\boldsymbol{\nu}_j = \mathbf{0}$ (respectively, $\bar{\boldsymbol{\nu}}_j = 0$), 
then the reactant (respectively, product) of the $j$-reaction
is the zero species, denoted by $\varnothing$, 
representing species that are not explicitly modelled.
We denote two irreversible reactions 
$(\sum_{l = 1}^N \nu_{i, l} X_l   \xrightarrow[]{k_{i}}  
\sum_{l = 1}^N \bar{\nu}_{i, l} X_l) \in \mathcal{R}$
and
$(\sum_{l = 1}^N \bar{\nu}_{i, l} X_l  \xrightarrow[]{k_{j}} 
\sum_{l = 1}^N \nu_{i, l} X_l) \in \mathcal{R}$
 jointly as the single reversible reaction
$(\sum_{l = 1}^N \nu_{i, l} X_l \xrightleftharpoons[k_j]{k_i} 
\sum_{l = 1}^N \bar{\nu}_{i, l} X_l) \in\mathcal{R}$.
The \emph{order of j-reaction} from network $\mathcal{R}$
is given by $\langle \mathbf{1}, \boldsymbol{\nu}_{j} \rangle \in \mathbb{Z}_{\ge}$.
The \emph{order of reaction network} $\mathcal{R}$ is given by the order of
its highest-order reaction; $\mathcal{R}$ of order higher than two
is said to be a \emph{higher-order} network.

\subsection{Stochastic model of reaction networks} \label{app:stochastic_model}
A suitable stochastic model for the time-evolution of discrete species copy-numbers
 $\mathbf{X}(t) = (X_1(t), \ldots,$ $X_N(t)) \in \mathbb{Z}_{\ge}^{N}$, 
where $t \in \mathbb{R}_{\ge}$ is the time-variable,
is a continuous-time discrete-space Markov chain~\cite{GillespieDerivation}.
The underlying probability mass function (PMF)
satisfies a partial difference-differential equation, 
called the \emph{chemical master equation} (CME)~\cite{RadekBook,VanKampen}, given by
\begin{align}
\frac{\partial}{\partial t} p(\mathbf{x},t)   =  \mathcal{L} p(\mathbf{x},t) & = 
\sum_{j = 1}^M (E_{\mathbf{x}}^{-\Delta \mathbf{\mathbf{x}}_j} - 1)
 \big(\alpha_j(\mathbf{x}) p(\mathbf{x},t) \big), \label{eq:stochasticmodel}
\end{align}
where $p(\mathbf{x},t)$ is the PMF,
i.e. the probability that the copy-number vector $\mathbf{X} = \mathbf{X}(t) 
\in \mathbb{Z}_{\ge}^N$ at time $t > 0$ is given by $\mathbf{x} 
\in \mathbb{Z}_{\ge}^N$. 
Here, the linear operator $\mathcal{L}$ is called the \emph{forward operator}, 
while the \emph{step operator} $E_{\mathbf{x}}^{-\Delta \mathbf{\mathbf{x}}} = 
\prod_{i = 1}^N E_{x_i}^{-\Delta x_i}$ is such that 
$E_{\mathbf{x}}^{-\Delta \mathbf{\mathbf{x}}} p(\mathbf{x},t)
 = p(\mathbf{x} - \Delta \mathbf{x},t)$.
Vector $\Delta \mathbf{x}_j = (\boldsymbol{\bar{\nu}}_j - \boldsymbol{\nu}_j) \in \mathbb{Z}^N$
is the \emph{reaction vector} of the $j$-reaction. 
 Function $\alpha_j(\mathbf{x})$ is the \emph{propensity function} of the $j$-reaction, 
and is given by
\begin{align}
\alpha_{j}(\mathbf{x}) & = k_{j} \mathbf{x}^{\underline{\boldsymbol{\nu}_j}} 
\equiv k_j \prod_{i = 1}^N x_i^{\underline{\nu_{j i}}}, \label{eq:dMAK}
\end{align}
where $x_i^{\underline{\nu_{j i}}} = x_i (x_i - 1) (x_i - 2) \ldots (x_i - \nu_{j i} + 1)$, 
with the convention that $x_i^{\underline{0}} \equiv 1$ for all $x_i \in \mathbb{Z}_{\ge}$.

The $l^2$-adjoint operator of $\mathcal{L}$,
denoted by $\mathcal{L}^*$ and called the \emph{backward operator}~\cite{PavliotisMultiscale}, 
is given by
\begin{align}
\mathcal{L}^{*} q(\mathbf{x}) & =  
\sum_{j = 1}^M \alpha_j(\mathbf{x}) (E_{\mathbf{x}}^{+\Delta \mathbf{\mathbf{x}}_j} - 1) 
q(\mathbf{x}), \label{eq:backwardoperator}
\end{align}
for a suitable class of functions $q(\mathbf{x}) : \mathbb{Z}_{\ge}^N \to \mathbb{R}$.
Null-space of an operator $\mathcal{L}^*$ is denoted by
$\mathcal{N}(\mathcal{L}^*) = \{q(\cdot) : \mathbb{Z}_{\ge}
 \to \mathbb{R} | \mathcal{L}^* q(\mathbf{x}) = 0\}$.

\section{Formal perturbation analysis of network~(\ref{eq:fulln})--(\ref{eq:subnetworksn})} \label{app:formal}
Let $\mathbf{x} = (x_1, x_2, \ldots, x_m, x_{m+1}, \ldots, x_N)
 \in \mathbb{Z}_{\ge}^{N}$ be the vector of copy-numbers for the species 
$\mathcal{X} = \{X_1, X_2, \ldots, X_m, X_{m+1}, \ldots, X_N \}$, 
and $\mathbf{y} = (y_1, y_2, \ldots, y_{n-2}) \in \mathbb{Z}_{\ge}^{n-2}$ 
the copy-number vector for the auxiliary species $\mathcal{Y} = 
\{Y_1, Y_2, \ldots, Y_{n-2}\}$ from the network~(\ref{eq:fulln})--(\ref{eq:subnetworksn}). 
Let us introduce new variables $\mathbf{\bar{x}} = 
(\bar{x}_1, \bar{x}_2, \ldots, \bar{x}_m, x_{m+1}, \ldots, x_N) \in \mathbb{Z}_{\ge}^{N}$ as follows:
 if there is only one distinct reactant species in the input 
network~(\ref{eq:reducedn}), then 
\begin{align}
\bar{x}_1 & = x_1 + \sum_{i = 1}^{n-2} (i+1) y_i, 
\; \; \; \textrm{if } m = 1. \label{eq:newvariablesn1}
\end{align}
On the other hand, if $m \ge 2$, then
\begin{align}
\bar{x}_l & = \begin{cases}
x_1 + \sum_{i=1}^{\nu_1} i y_{i} + \sum_{i=\nu_1 + 1}^{n-2} \nu_1 y_i, & \text{if } l = 1, \\
x_2 + \sum_{i=1}^{\nu_1} y_{i} + \sum_{i=\nu_1 + 1}^{\nu_1 + \nu_2 -  1} (i - \nu_1 + 1) y_i +
 \sum_{i = \nu_1 + \nu_2}^{n-2} \nu_2 y_i, & \text{if } l = 2, \text{ and } m \ne 2, \\  
x_l + \sum_{i= \sum_{j=1}^{l-1} \nu_j}^{-1 + \sum_{j=1}^{l} \nu_j} (1 + i - \sum_{j=1}^{l-1} \nu_j) y_i 
+ \sum_{i= \sum_{j=1}^{l} \nu_j}^{n-2} \nu_l y_i, & \text{if } l \in \{3, 4, \ldots, m-1 \}, \\
x_m + \delta_{m,2} \sum_{i=1}^{\nu_1} y_{i} + 
\sum_{i=  \delta_{m,2} + \sum_{j=1}^{m-1} \nu_j}^{n - 2} (1 + i - \sum_{j=1}^{m-1} \nu_j) y_i, 
& \text{if } l = m.
\end{cases} \label{eq:newvariablesn}
\end{align}
In what follows, we define the faster network
\begin{align}
\mathcal{R}_{0}^{\varepsilon}: \; \; 
Y_{1} &\xrightarrow[]{\frac{1}{\varepsilon}}  X_l + X_r, \nonumber \\
Y_{i} &\xrightarrow[]{\frac{1}{\varepsilon}}  X_l + Y_{i - 1}, 
\; \; \; \textrm{for all } i \in \{2, 3, \ldots, n-2\}.
\label{eq:fastsubnetworksn}
\end{align}

The CME corresponding to~(\ref{eq:fulln})--(\ref{eq:subnetworksn}), 
rescaled in time according to $t = \tau/\varepsilon^{n-2}$, and expressed 
in terms of the new variables $\mathbf{\bar{x}}$, is given by
\begin{align}
\frac{\mathrm{d}}{\mathrm{d} \tau} p_{\varepsilon}(\mathbf{\bar{x}},\mathbf{y},\tau) & = 
\left(\frac{1}{\varepsilon^{n-1}}\mathcal{L}_0  + 
\frac{1}{\varepsilon^{n-2}}  \sum_{i = 1}^{n-1} \mathcal{L}_{i} \right)  
p_{\varepsilon}(\mathbf{\bar{x}},\mathbf{y},\tau), \label{eq:rescaledCMEn}
\end{align}
where $\mathcal{L}_0$ is the forward operator of network $\mathcal{R}_0^1$
in the new coordinates, given by
\begin{align}
\mathcal{L}_0  & = \sum_{i = 1}^{n-2} \mathcal{L}_0^{i}, \; \; \; \; \; \; 
\mathcal{L}_0^i = \left(E_{y_{i-1}}^{-1} E_{y_{i}}^{+1} - 1 \right) 
y_i, \; \; \textrm{for all } i \in  \{1, 2, \ldots, n-2 \}, \label{eq:operatorsn0}
\end{align}
while $\sum_{i = 1}^{n-1} \mathcal{L}_{i}$ is the forward operator
of the slower sub-network 
$\mathcal{R}_{\varepsilon} \setminus \mathcal{R}_{0}^{\varepsilon}$, with
\begin{align}
\mathcal{L}_i & = \begin{cases}
\left(E_{y_{i-1}}^{+1} E_{y_{i}}^{-1} - 1 \right) 
y_{i-1} \alpha_i(\mathbf{\bar{x}},\mathbf{y}),  & \textrm{if } i \in \{1, 2, \ldots, n-2 \}, \\
 \left(E_{\mathbf{\bar{x}}}^{-\Delta \mathbf{\bar{x}}} 
E_{\mathbf{y}}^{-\Delta \mathbf{y}} - 1 \right) y_{i-1} \alpha_{i}(\mathbf{\bar{x}}, \mathbf{y}),
& \textrm{if } i = (n-1).
\end{cases}\label{eq:operatorsn1}
\end{align}
Here, we take the convention that $y_0 \equiv 1$, and that the 
operators $E_{y_{0}}^{\pm 1}$ are the identity operator.
Function $y_{i-1} \alpha_i(\mathbf{\bar{x}},\mathbf{y})$ is
 the propensity function of the forward reaction in the sub-network
 $\mathcal{R}_{i}^{\varepsilon}$ from~(\ref{eq:subnetworksn}), expressed in terms of the 
new variables~(\ref{eq:newvariablesn1})--(\ref{eq:newvariablesn}). 
Let us note that $\mathcal{L}_{n-1}$ is the only operator which 
acts on the variable $\mathbf{\bar{x}}$, while the rest of the operators 
act only on $\mathbf{y}$. On the other hand, it follows from~(\ref{eq:subnetworksn})  that 
the reaction vector $\Delta \mathbf{y} = (\Delta y_1, \Delta y_2, \ldots, \Delta y_{n-3}, \Delta y_{n-2}) = 
(\bar{\gamma}_1, \bar{\gamma}_2, \ldots, \bar{\gamma}_{n-3}, \bar{\gamma}_{n-2} - 1)$,
while the reaction vector $\Delta \mathbf{\bar{x}}$ is obtained
by formally applying the difference operator 
$\Delta$ on~(\ref{eq:newvariablesn1})--(\ref{eq:newvariablesn}). 

Let us write the solution of~(\ref{eq:rescaledCMEn}) as the perturbation series
\begin{align}
p_{\varepsilon}(\mathbf{\bar{x}},\mathbf{y},\tau)  = 
\sum_{i = 0}^{n-1} \varepsilon^i p_i(\mathbf{\bar{x}},\mathbf{y},\tau)  + \ldots. 
\label{eq:powerseriesn}
\end{align}
Substituting~(\ref{eq:powerseriesn}) into~(\ref{eq:rescaledCMEn}), 
and equating terms of equal powers in $\varepsilon$, the following 
system of $n$ equations is obtained:
\begin{align}
\mathcal{O} \left(\frac{1}{\varepsilon^{n-i}} \right): \;  
\mathcal{L}_{0} p_{i-1}(\mathbf{\bar{x}},\mathbf{y},\tau) & = 
- \sum_{i = 1}^{n-1} \mathcal{L}_{i} p_{i-2}(\mathbf{\bar{x}},\mathbf{y},\tau), \, \,  \, \, \textrm{for all } i \in \{1, 2, \ldots, n-1 \}, \nonumber \\
\mathcal{O}(1): \;  \mathcal{L}_{0} p_{n-1}(\mathbf{\bar{x}},\mathbf{y},\tau)  & = 
\frac{\mathrm{d}}{\mathrm{d} \tau} p_{0}(\mathbf{\bar{x}},\mathbf{y},\tau)
- \sum_{i = 1}^{n-1} \mathcal{L}_{i} p_{n-2}(\mathbf{\bar{x}},\mathbf{y},\tau), \label{eq:QSAn}
\end{align}
with the convention that $p_{-1}(\mathbf{\bar{x}},\mathbf{y},\tau) \equiv 0$.

\emph{Order $1/\varepsilon^{n-1}$ equation}. 
Operator $\mathcal{L}_0$, given in~(\ref{eq:operatorsn0}), acts and depends only on the 
variable $\mathbf{y}$, and each summand $\mathcal{L}_0^{i}$ is multiplied 
on the right by a factor $y_i$; therefore
\begin{align}
p_0(\mathbf{\bar{x}},\mathbf{y},\tau) & = p_0(\mathbf{\bar{x}},\tau) 
\prod_{i=1}^{n-2} \delta_{y_i,0}, \; \; \; \; \; 
\sum_{\mathbf{\bar{x}}} p_0(\mathbf{\bar{x}},\tau) = 1, 
\; \; \textrm{for all } \tau \ge 0. \label{eq:solp0n}
\end{align}

\emph{Order $1/\varepsilon^{n-2}$ equation}. Since each of the operators
$\{\mathcal{L}_{i}\}_{i = 2}^{n-1}$ from~(\ref{eq:operatorsn1}) is 
multiplied on the right by a nonconstant factor $y_{i-1}$, equation~(\ref{eq:solp0n}) 
implies
\begin{align}
\left( \sum_{i = 1}^{n-1} \mathcal{L}_{i} \right) p_0(\mathbf{\bar{x}},\mathbf{y},\tau) & = 
p_0(\mathbf{\bar{x}},\tau) \left( \mathcal{L}_1 
\delta_{y_1,0} \right) \prod_{i=2}^{n-2} \delta_{y_i,0}. \label{eq:solv1}
\end{align}
The null-space of the backward operator is given by
$\mathcal{N}(\mathcal{L}_0^*) = \{1\}$,
and the solvability condition is $0 = \langle 1, 
\sum_{i = 1}^{n-1} \mathcal{L}_{i} p_0(\mathbf{\bar{x}},\mathbf{y},\tau)\rangle_{\mathbf{y}} = 
\langle \mathcal{L}_1^* 1, 
p_0(\mathbf{\bar{x}},\tau) \prod_{i=1}^{n-2} \delta_{y_i,0}\rangle_{\mathbf{y}}$
is identically satisfied.

Let us write the solution of the $\mathcal{O}(1/\varepsilon^{n-2})$
 equation from~(\ref{eq:QSAn}) in the separable form
\begin{align}
p_1(\mathbf{\bar{x}},\mathbf{y},\tau) & = p_0(\mathbf{\bar{x}},\tau) 
p_1(y_1; \, \mathbf{\bar{x}}) \prod_{i=2}^{n-2} \delta_{y_i, 0}, \label{eq:p1n}
\end{align}
so that
\begin{align}
\mathcal{L}_0 p_1(\mathbf{\bar{x}},\mathbf{y},\tau) & = 
p_0(\mathbf{\bar{x}},\tau) \prod_{i=2}^{n-2} \delta_{y_i, 0} \mathcal{L}_0^1 
p_1(y_1; \, \mathbf{\bar{x}}), \label{eq:p1nL0}
\end{align}
Substituting~(\ref{eq:solv1}) and~(\ref{eq:p1nL0}) into
the $\mathcal{O}(1/\varepsilon^{n-2})$ equation, 
and using the operator equality $(E_{y_1}^{-1} - 1) = 
- (E_{y_1}^{+1} - 1) E_{y_1}^{-1}$, one obtains
\begin{align}
 \prod_{i=2}^{n-2} \delta_{y_i, 0} (E_{y_1}^{+1} - 1) 
\left(y_1 p_1(y_1; \, \mathbf{\bar{x}}) - E_{y_1}^{-1} 
\alpha_1(\mathbf{\bar{x}},\mathbf{y}) \delta_{y_1, 0} \right) & = 0. 
\label{eq:solp1n1}
\end{align}
Equation~(\ref{eq:solp1n1}) is identically satisfied if 
$(y_2, y_3, \ldots, y_{n-2}) \ne \mathbf{0}_{n-3}$, where $\mathbf{0}_{n}$ 
is the zero element of $\mathbb{Z}_{\ge}^{n}$. 
On the other hand, if 
$(y_2, y_3, \ldots, y_{n-2}) = \mathbf{0}_{n-3}$, it follows that the solutions satisfy
\begin{align}
y_1 p_1(y_1; \, \mathbf{\bar{x}}) & = E_{y_1}^{-1} 
\alpha_1 \left(\mathbf{\bar{x}},(y_1, \mathbf{0}_{n-3}) \right) \delta_{y_1, 0}. 
\label{eq:solp1n}
\end{align}

\emph{Order $1/\varepsilon^{n-3}$ equation}. It follows from~(\ref{eq:operatorsn1}) 
and~(\ref{eq:p1n}) that
\begin{align}
\left(\sum_{i = 1}^{n-1} \mathcal{L}_{i} \right) p_1(\mathbf{\bar{x}},\mathbf{y},\tau) & = 
p_0(\mathbf{\bar{x}},t) \left( (\mathcal{L}_1 + \mathcal{L}_2)
p_1(y_1; \, \mathbf{\bar{x}}) \delta_{y_2, 0} \right) 
\prod_{i=3}^{n-2} \delta_{y_i, 0}, \label{eq:solv2}
\end{align}
and the solvability condition is identically satisfied.
Let us write the solution of the $\mathcal{O}(1/\varepsilon^{n-3})$ 
equation from~(\ref{eq:QSAn}) in the separable form
\begin{align}
p_2(\mathbf{\bar{x}},\mathbf{y},\tau) & = 
p_0(\mathbf{\bar{x}},\tau) p_2(y_1, y_2; \, \mathbf{\bar{x}}) 
\prod_{i=3}^{n-2} \delta_{y_i, 0}, \label{eq:p2n}
\end{align}
so that
\begin{align}
\mathcal{L}_0 p_2(\mathbf{\bar{x}},\mathbf{y},\tau) & = 
p_0(\mathbf{\bar{x}},\tau)  \prod_{i=3}^{n-2} \delta_{y_i, 0} 
\left(\mathcal{L}_0^1 + \mathcal{L}_0^2 \right) 
p_2(y_1, y_2; \, \mathbf{\bar{x}}). \label{eq:p2nL0}
\end{align}
Substituting~(\ref{eq:solv2}) and~(\ref{eq:p2nL0}) into
the $\mathcal{O}(1/\varepsilon^{n-3})$ equation, 
and using the operator equalities $(E_{y_1}^{-1} - 1) = - (E_{y_1}^{+1} - 1) E_{y_1}^{-1}$
and $(E_{y_1}^{+1} E_{y_2}^{-1} - 1) = - (E_{y_1}^{-1} E_{y_2}^{+1} - 1) 
E_{y_1}^{+1} E_{y_2}^{-1}$, one obtains
\begin{align}
0 & = \prod_{i=3}^{n-2} \delta_{y_i, 0} (E_{y_1}^{+1} - 1) 
\left[y_1 p_2(y_1, y_2; \, \mathbf{\bar{x}}) - E_{y_1}^{-1} 
\alpha_1(\mathbf{\bar{x}},\mathbf{y}) \delta_{y_2, 0} p_1(y_1; \, \mathbf{\bar{x}}) \right]
 \nonumber \\
& + \prod_{i=3}^{n-2} \delta_{y_i, 0} (E_{y_1}^{-1} E_{y_2}^{+1} - 1)
\left[y_2 p_2(y_1, y_2; \, \mathbf{\bar{x}}) -  E_{y_1}^{+1} E_{y_2}^{-1} 
\alpha_2(\mathbf{\bar{x}},\mathbf{y}) \delta_{y_2, 0} y_1 p_1(y_1; \, \mathbf{\bar{x}}) \right]. 
\label{eq:solp2n1}
\end{align} 
Equation~(\ref{eq:solp2n1}) is identically satisfied if 
$(y_3, y_4, \ldots, y_{n-2}) \ne \mathbf{0}_{n-4}$. 
On the other hand, if $(y_3, y_4, \ldots, y_{n-2}) = \mathbf{0}_{n-4}$, 
the solutions satisfy
\begin{align}
y_1 p_2(y_1, y_2; \, \mathbf{\bar{x}}) & = E_{y_1}^{-1} \alpha_1 
\left(\mathbf{\bar{x}},(y_1,y_2,\mathbf{0}_{n-4}) \right) 
\delta_{y_2, 0} p_1(y_1; \, \mathbf{\bar{x}}), \label{eq:solp2nauxiliary}
\end{align}
and
\begin{align}
y_2 p_2(y_1, y_2; \, \mathbf{\bar{x}}) & =  E_{y_1}^{+1} E_{y_2}^{-1} 
\alpha_2 \left(\mathbf{\bar{x}},(y_1,y_2,\mathbf{0}_{n-4}) \right) 
\delta_{y_2, 0} \left( y_1 p_1(y_1; \, \mathbf{\bar{x}}) \right) \nonumber \\
& = E_{y_1}^{+1} E_{y_2}^{-1} \alpha_2 
\left(\mathbf{\bar{x}},(y_1,y_2,\mathbf{0}_{n-4}) \right) 
\delta_{y_2, 0} \left( E_{y_1}^{-1} \alpha_1 
\left(\mathbf{\bar{x}},(y_1, \mathbf{0}_{n-3}) \right) \delta_{y_1, 0} \right) \nonumber \\
& = \delta_{y_1, 0} \alpha_1 \left(\mathbf{\bar{x}},(y_1, \mathbf{0}_{n-3}) \right) 
 E_{y_1}^{+1} E_{y_2}^{-1} 
\alpha_2 \left(\mathbf{\bar{x}},(y_1,y_2,\mathbf{0}_{n-4}) \right) 
\delta_{y_2, 0}, \label{eq:solp2n}
\end{align}
where we have used~(\ref{eq:solp1n}) when going from the first
to the second line in~(\ref{eq:solp2n}). 

\emph{Order $1/\varepsilon^{n-i}$ equation}, $i \in \{3, 4, \ldots, n-1 \}$. One can inductively
proceed to the higher-order equations from~(\ref{eq:QSAn}), 
with the solutions of the $\mathcal{O}(1/\varepsilon^{n-i})$ equation written in the separable form
\begin{align}
p_{i-1}(\mathbf{\bar{x}},\mathbf{y},t) & = p_0(\mathbf{\bar{x}},t) 
p_{i-1}(y_1, \ldots, y_{i-1}; \, \mathbf{\bar{x}}) \prod_{j=i}^{n-2} \delta_{y_j, 0}, 
\; \; \; \textrm{for all } i \in \{1, 2, \ldots, n-1 \}, \label{eq:pin}
\end{align}
with the convention that $\prod_{i = a}^{b} f(i) = 1$ if $a > b$, where $f(i)$ is an arbitrary function of $i$, 
and $p_{0}(y_{0}; \, \mathbf{\bar{x}}) \equiv 1$ (see also equations~(\ref{eq:solp0n}), (\ref{eq:p1n}) and~(\ref{eq:p2n})). 
It can be readily shown that the results~(\ref{eq:solp1n}) and~(\ref{eq:solp2n})
generalize to
\begin{align}
y_{i} p_{i}(y_1, \ldots, y_i; \, \mathbf{\bar{x}}) & = 
\left( \prod_{j = 1}^{i-1} \delta_{y_{j}, 0} E_{y_{j-1}}^{+1} 
\alpha_j \Big(\mathbf{\bar{x}}, (y_1, \ldots, y_j, \mathbf{0}_{n-2-j}) \Big) \right) \nonumber \\
& \times E_{y_{i-1}}^{+1} E_{y_{i}}^{-1} \alpha_{i} \Big(\mathbf{\bar{x}}, 
(y_1, \ldots, y_i, \mathbf{0}_{n-2-i}) \Big) \delta_{y_{i}, 0}, 
\; \; \; \textrm{for all } i \in \{1, 2, \ldots, n-2 \}. \label{eq:solpnn}
\end{align}
As is shortly shown, 
the effective CME, at the order one, depends on $p_{n-2}$
only via the product $y_{n-2} p_{n-2}$  satisfying~(\ref{eq:solpnn}); 
see equation~(\ref{eq:preeffectiven2}).

\emph{Order $1$ equation}. The solvability condition 
is given by $0 = \mathrm{d}/\mathrm{d} \tau p_{0}(\mathbf{\bar{x}},\tau)
- \langle 1, (\sum_{i = 1}^{n-1} \mathcal{L}_{i}) p_{n-2}(\mathbf{\bar{x}},\mathbf{y},\tau) \rangle_{\mathbf{y}}$.
It follows from~(\ref{eq:operatorsn1}) that $(\sum_{i = 1}^{n-2} \mathcal{L}_i)^* 
1 = 0$, implying that
$\langle 1, \sum_{i = 1}^{n-1} \mathcal{L}_{i}
p_{n-2}(\mathbf{\bar{x}},\mathbf{y},\tau)\rangle_{\mathbf{y}}
= \langle 1, \mathcal{L}_{n-1} 
p_{n-2}(\mathbf{\bar{x}},\mathbf{y},\tau)\rangle_{\mathbf{y}} $;
therefore, solvability condition becomes
\begin{align}
\frac{\mathrm{d}}{\mathrm{d} \tau}  p_{0}(\mathbf{\bar{x}},\tau) & = 
\left \langle 1, \mathcal{L}_{n-1} p_{n-2}(\mathbf{\bar{x}},\mathbf{y},\tau) \right \rangle_{\mathbf{y}}. 
 \label{eq:preeffectiven}
\end{align}
Let us simplify the right-hand side from~(\ref{eq:preeffectiven}):
\begin{align}
\left \langle 1, \mathcal{L}_{n-1} p_{n-2}(\mathbf{\bar{x}},\mathbf{y},\tau) \right \rangle_{\mathbf{y}}
& = \left \langle 1, \left(E_{\mathbf{\bar{x}}}^{-\Delta \mathbf{\bar{x}}} 
E_{\mathbf{y}}^{-\Delta \mathbf{y}} - 1 \right) \alpha_{n-1}(\mathbf{\bar{x}}, \mathbf{y}) 
y_{n-2} p_{n-2}(\mathbf{\bar{x}},\mathbf{y},\tau) \right\rangle_{\mathbf{y}} \nonumber \\
& = \left \langle E_{\mathbf{y}}^{+\Delta \mathbf{y}} 1, 
E_{\mathbf{\bar{x}}}^{-\Delta \mathbf{\bar{x}}} \alpha_{n-1}(\mathbf{\bar{x}}, \mathbf{y}) 
y_{n-2} p_{n-2}(\mathbf{\bar{x}},\mathbf{y},\tau) \right \rangle_{\mathbf{y}} \nonumber \\
& - \left \langle 1, \alpha_{n-1}(\mathbf{\bar{x}}, \mathbf{y}) 
y_{n-2} p_{n-2}(\mathbf{\bar{x}},\mathbf{y},\tau) \right \rangle_{\mathbf{y}} \nonumber \\
& = \left \langle 1, \left(E_{\mathbf{\bar{x}}}^{-\Delta \mathbf{\bar{x}}} - 1 \right) 
\alpha_{n-1}(\mathbf{\bar{x}}, \mathbf{y}) 
y_{n-2} p_{n-2}(\mathbf{\bar{x}},\mathbf{y},\tau) \right \rangle_{\mathbf{y}} \nonumber \\
& = \left(E_{\mathbf{\bar{x}}}^{-\Delta \mathbf{\bar{x}}} - 1 \right) p_{0}(\mathbf{\bar{x}},\tau)
\left \langle  \alpha_{n-1}(\mathbf{\bar{x}}, \mathbf{y}), 
y_{n-2} p_{n-2}(\mathbf{y}; \, \mathbf{\bar{x}}) \right \rangle_{\mathbf{y}}, \label{eq:preeffectiven2}
\end{align}
where, when going to the last line, we use $p_{n-2}(\mathbf{\bar{x}},\mathbf{y},\tau)
= p_{0}(\mathbf{\bar{x}},\tau) p_{n-2}(\mathbf{y}; \, \mathbf{\bar{x}})$.
Hence, the effective CME depends on the $p_{n-2}$
only via the product $y_{n-2} p_{n-2}$.
Using~(\ref{eq:solpnn}) with  $i = n-2$, one obtains
\begin{align}
\left \langle \alpha_{n-1}(\mathbf{\bar{x}}, \mathbf{y}),
y_{n-2} p_{n-2}(\mathbf{y}; \, \mathbf{\bar{x}}) \right\rangle_{\mathbf{y}} 
& = \langle \alpha_{n-1}(\mathbf{\bar{x}}, \mathbf{y})  
\left( \prod_{j = 1}^{n-3} \delta_{y_{j}, 0} E_{y_{j-1}}^{+1} 
\alpha_j \left(\mathbf{\bar{x}}, (y_1, \ldots, y_j, \mathbf{0}_{n-2-j})\right) \right), \nonumber \\
& E_{y_{n-3}}^{+1} E_{y_{n-2}}^{-1} \alpha_{n-2}
 (\mathbf{\bar{x}},\mathbf{y}) \delta_{y_{n-2}, 0} \rangle_{\mathbf{y}} \nonumber \\
& = \langle 1,  \left( \prod_{j = 1}^{n-2} \delta_{y_{j}, 0} \right)
 \left( \prod_{j = 1}^{n-3} E_{y_{j-1}}^{+1} 
\alpha_j \left(\mathbf{\bar{x}}, (y_1, \ldots, y_j, \mathbf{0}_{n-2-j})\right) \right) \nonumber \\
& \left( E_{y_{n-3}}^{+1} \alpha_{n-2} (\mathbf{\bar{x}}, \mathbf{y} \right)
\left( E_{y_{n-2}}^{+1} \alpha_{n-1}(\mathbf{\bar{x}}, \mathbf{y} \right) \rangle_{\mathbf{y}}  \nonumber \\
& =  \prod_{j = 1}^{n-1} \alpha_j 
\left(\mathbf{\bar{x}}, (\mathbf{0}_{j-2}, 1, \mathbf{0}_{n - (j+1)})\right),
\label{eq:propensityeffectiven}
\end{align}
with the convention that $\alpha_1 (\mathbf{\bar{x}}, (\mathbf{0}_{-1}, 1, \mathbf{0}_{n -2})) 
\equiv \alpha_1 (\mathbf{\bar{x}}, \mathbf{0}_{n -2})$, $\alpha_2 (\mathbf{\bar{x}}, 
(\mathbf{0}_{0}, 1, \mathbf{0}_{n -3}))$ $\equiv \alpha_2 (\mathbf{\bar{x}}, 
(1,\mathbf{0}_{n -3}))$, and $\alpha_{n-1} (\mathbf{\bar{x}}, 
(\mathbf{0}_{n-3}, 1, \mathbf{0}_{0})) \equiv \alpha_{n-1} 
(\mathbf{\bar{x}}, (\mathbf{0}_{n-3}, 1))$. 
In words, propensity function $\alpha_j$ is evaluated at $y_{j-1} = 1$,
and $y_i = 0$ for $i \ne (j-1)$ in~(\ref{eq:propensityeffectiven}).

If $m = 1$, using~(\ref{eq:newvariablesn1}), it follows that 
$\alpha_1(x_1, \mathbf{0}_{n -2}) = \kappa_1 \bar{x}_1 (\bar{x}_1 - 1)$, 
and $\alpha_j \left(\bar{x}_1, (\mathbf{0}_{j-2}, 1, \mathbf{0}_{n - (j+1)})\right)
= \kappa_j (\bar{x}_1 - j)$, for $j \in \{2, 3, \ldots, n-1 \}$. Hence, in this case, 
\begin{align}
\prod_{j = 1}^{n-1} \alpha_j \left(\mathbf{\bar{x}}, 
(\mathbf{0}_{j-2}, 1, \mathbf{0}_{n - (j+1)})\right)& = 
\left( \prod_{j = 1}^{n-1} \kappa_j \right) \bar{x}_1^{\underline{n}}, 
\; \; \; \text{if } m = 1. \label{eq:effectiveprop1}
\end{align}
Similarly, in the case $m \ge 2$, using~(\ref{eq:newvariablesn}), 
one obtains 
\begin{align}
\prod_{j = 1}^{n-1} \alpha_j \left(\mathbf{\bar{x}}, 
(\mathbf{0}_{j-2}, 1, \mathbf{0}_{n - (j+1)})\right)& = 
\left( \prod_{j = 1}^{n-1} \kappa_j \right) \prod_{l = 1}^m \bar{x}_l^{\underline{\nu_l}}, 
\; \; \; \text{if } m \ge 2. \label{eq:effectivepropn}
\end{align}
Substituting~(\ref{eq:preeffectiven2})--(\ref{eq:effectivepropn}) 
into~(\ref{eq:preeffectiven}), and changing the time back to
the original scale, $\tau = \varepsilon^{n-2} t$, one obtains
the effective CME
\begin{align}
 \frac{\mathrm{d}}{\mathrm{d} t}  p_0(\mathbf{\bar{x}},t)  & = 
\left(E_{\mathbf{\bar{x}}}^{-\Delta \bar{\mathbf{x}}} - 1 \right) 
\left(\varepsilon^{n-2}  \prod_{j = 1}^{n-1} \kappa_j  \right)
\prod_{l = 1}^m \bar{x}_l^{\underline{\nu_l}} p_0(\mathbf{\bar{x}},t). 
\label{eq:effectiveCMEn}
\end{align}
Assuming convergence,
it follows from~(\ref{eq:solp0n}) that 
the time-dependent copy-number vector 
$\mathbf{Y}(t) \in \mathbb{Z}_{\ge}^{n-2}$ converges
weakly (in distribution) to $\mathbf{0} \in \mathbb{Z}_{\ge}^{n-2}$ 
(a deterministic variable), and hence it also converges to zero
in probability. It then follows from~(\ref{eq:newvariablesn1})--(\ref{eq:newvariablesn})
that $\bar{\mathbf{X}}(t)$  converges to $\mathbf{X}(t)$ in probability
as $\varepsilon \to 0$, ensuring that~(\ref{eq:effectiveCMEn}) tracks copy-numbers
of the species $\mathcal{X}$. 

\subsubsection*{Kinetic and stoichiometric conditions}
In order for the effective CME~(\ref{eq:effectiveCMEn})
to match the CME of the
input network~\rm (\ref{eq:reducedn}), 
the kinetic condition~(\ref{eq:kineticn}) must hold. 
Furthermore, we require that $\Delta \bar{\mathbf{x}} = \Delta \mathbf{x} = 
(\boldsymbol{\bar{\nu}} - \boldsymbol{\nu})$.
In the special case when $(\bar{\gamma}_1, \bar{\gamma}_2, \ldots, \bar{\gamma}_{n-3}) =  
(0,0,\ldots,0)$, applying the difference operator $\Delta$ on~(\ref{eq:newvariablesn}), 
and setting $\Delta \bar{x}_l = (\bar{\nu}_l - \nu_l)$, one obtains 
\begin{align}
\bar{\nu}_l - \nu_l & = \Delta x_l   + 
(\nu_l - \delta_{l,m}) \Delta y_{n-2} \nonumber \\
& = (\tilde{\nu}_l -\delta_{l,m}) + 
(\nu_l - \delta_{l,m}) (\bar{\gamma}_{n-2} - 1),
\; \; \; \textrm{for all } l \in \{1, 2, \ldots, m\}, 
\label{eq:stoichiometryn}
\end{align}
which implies the stoichiometric
conditions~(\ref{eq:stoichiometricn}).

\section{Proof of Theorem~\ref{theorem:convergencen}} \label{app:convergence}
Consider the output network~{\rm (\ref{eq:fulln})}--{\rm (\ref{eq:subnetworksn})}
under the kinetic scaling~(\ref{eq:symmetrickinetic}), with the
the CME~(\ref{eq:rescaledCMEn}) in the original time $t$ given by
\begin{align}
\frac{\mathrm{d}}{\mathrm{d} t} p_{\varepsilon}(\mathbf{\bar{x}},\mathbf{y},t) &  
= \mathcal{L}_{\varepsilon}^n p_{\varepsilon}(\mathbf{\bar{x}},\mathbf{y},t)
= \left(\frac{1}{\varepsilon} \mathcal{L}_0  + 
\frac{1}{\varepsilon^{\frac{n-2}{n-1}}} \sum_{i = 1}^{n-1} \mathcal{L}_{i} \right)  
p_{\varepsilon}(\mathbf{\bar{x}},\mathbf{y},t). \label{eq:CMEn}
\end{align}
Substituting into~(\ref{eq:CMEn}) the perturbation series
\begin{align}
p_{\varepsilon}(\mathbf{\bar{x}},\mathbf{y},t)  = 
\sum_{i = 0}^{n-1} \varepsilon^{\frac{i}{n-1}} p_i(\mathbf{\bar{x}},\mathbf{y},t)  + 
\ldots, \label{eq:perturbationn_scaledproof}
\end{align}
one obtains
\begin{align}
\mathcal{O} \left(\frac{1}{\varepsilon^{1 - \frac{i}{n-1}}} \right): \;  
\mathcal{L}_{0} p_{i}(\mathbf{\bar{x}},\mathbf{y},t) & = 
- \left(\sum_{i = 1}^{n-1} \mathcal{L}_{i}\right) p_{i-1}(\mathbf{\bar{x}},\mathbf{y},t), \, \,  \, \, \textrm{for all } i \in \{0, 1, \ldots, n-2 \}, \nonumber \\
\mathcal{O}(1): \;  \mathcal{L}_{0} p_{n-1}(\mathbf{\bar{x}},\mathbf{y},t)  & = 
\frac{\mathrm{d}}{\mathrm{d} t} p_{0}(\mathbf{\bar{x}},\mathbf{y},t)
-  \left(\sum_{i = 1}^{n-1} \mathcal{L}_{i}\right) p_{n-2}(\mathbf{\bar{x}},\mathbf{y},t). \label{eq:QSAn_scaled}
\end{align}
System~(\ref{eq:QSAn_scaled}) has the same form
as~(\ref{eq:QSAn}) and, therefore, the same solutions
as those from Section~\ref{app:formal}. In particular, the 
zero-order term is given~(\ref{eq:solnp0_scaled})--(\ref{eq:effectiveCMEn_scaled}).

\begin{proof}
By construction from Section~\ref{app:formal}, there exist functions 
$\{p_i(\bar{\mathbf{x}},\mathbf{y},t)\}_{i = 1}^{n-1}$
such that system~(\ref{eq:QSAn_scaled}) is satisfied. 
Writing $p_i(t) = p_i(\bar{\mathbf{x}},\mathbf{y},t)$ for all $i \in \{0, 1, \ldots, n-1\}$,
we define a remainder function $r_{\varepsilon}^n(t) = r_{\varepsilon}^n(\bar{\mathbf{x}},\mathbf{y},t)$ via
\begin{align}
p_{\varepsilon}(t) & = 
\sum_{i = 0}^{n-1} \varepsilon^{\frac{i}{n-1}} p_i(t) + r_{\varepsilon}^n(t).
 \label{eq:remaindern}
\end{align}
Substituting~(\ref{eq:remaindern}) into~(\ref{eq:CMEn}), and using~(\ref{eq:QSAn_scaled}) 
together with $p_{\varepsilon}(0) = p_0(0)$, 
one obtains an initial-value problem for the remainder:
\begin{align}
\frac{\mathrm{d}}{\mathrm{d} t} r_{\varepsilon}^n(t) - \mathcal{L}_{\varepsilon}^n r_{\varepsilon}^n(t) & = 
\varepsilon^{\frac{1}{n-1}} \left(- \frac{\mathrm{d}}{\mathrm{d} t} p_1(t) + \left( \sum_{i = 1}^{n-1} \mathcal{L}_{i} \right) p_{n-1}(t) \right) + 
\sum_{i = 2}^{n-1} \varepsilon^{\frac{i}{n-1}} \frac{\mathrm{d}}{\mathrm{d} t} p_i(t), 
\; \; \; 
r_{\varepsilon}^n(0) = - \sum_{i = 1}^{n-1} \varepsilon^{\frac{i}{n-1}} p_i(0).
\label{eq:errorn}
\end{align}
Solving~(\ref{eq:errorn}), applying the $l^1$-norm on
$\mathbb{S} \subset \mathbb{Z}_{\ge}^{N + (n-2)}$, the triangle inequality, 
and using the fact that $\| e^{\mathcal{L}_{\varepsilon}^n t} \|_{l_1(\mathbb{S})} \le 1$, one obtains
\begin{align}
 \| r_{n} (t) \|_{l_1(\mathbb{S})} & \le  \varepsilon^{\frac{1}{n-1}} \left[ \| p_1(0) \|_{l_1(\mathbb{S})}
+ t  \, \textrm{sup}_{0 \le s \le t} \left(\left \|\frac{\mathrm{d}}{\mathrm{d} s} p_1(s) \right\|_{l_1(\mathbb{S})} + 
\left(\sum_{i = 1}^{n-1} \left \|\mathcal{L}_{i} \right\|_{l_1(\mathbb{S})} \right) \|p_{n-1}(s)\|_{l_1(\mathbb{S})} \right)
\right] \nonumber \\
& + \sum_{i = 2}^{n-1} \varepsilon^{\frac{i}{n-1}} \left[ \|p_i(0) \|_{l_1(\mathbb{S})}
+ t \, \textrm{sup}_{0 \le s \le t} \left\| \frac{\mathrm{d}}{\mathrm{d} s} p_i(s) \right\|_{l_1(\mathbb{S})} \right].
\label{eq:errorboundn}
\end{align}
It follows from~(\ref{eq:QSAn_scaled}) that there exist
functions $\{p_i(\bar{\mathbf{x}},\mathbf{y},t)\}_{i = 1}^{n-1}$ that are bounded
and have bounded time-derivatives.
Therefore, $\| r_{\varepsilon} (t) \|_{l_1(\mathbb{S})} = \mathcal{O}(\varepsilon^{1/(n-1)})$
as $\varepsilon \to 0$ for all $t \in [0,T]$ which, together 
with~(\ref{eq:remaindern}), implies~(\ref{eq:convergencen}).
\end{proof}

\label{lastpage}


\begin{thebibliography}{9}
\bibitem{Feinberg} Feinberg, M. \textit{Lectures on chemical reaction networks}. 
Delivered at the Mathematics Research Center, University of Wisconsin, 1979.

\bibitem{Janos} \'{E}rdi, P., T\'{o}th, J. 
\textit{Mathematical models of chemical reactions. Theory and applications of deterministic and stochastic Models}. 
Manchester University Press, Princeton University Press, 1989.

\bibitem{SysBio1} 
Vilar, J. M. G., Kueh, H. Y., Barkai, N., Leibler, S., 2002.
Mechanisms of noise-resistance in genetic oscillators.
\textit{Proceedings of the National Academy of Sciences of the United States of America}, 
\textbf{99}(\textbf{9}): 5988--5992.

\bibitem{SysBio2} 
Dublanche, Y., Michalodimitrakis, K., Kummerer, N., Foglierini, M., Serrano, L., 2006.
Noise in transcription negative feedback loops: simulation and experimental analysis.
\textit{Molecular Systems Biology}, \textbf{2}(\textbf{41}): E1--E12.

\bibitem{SysBio3} Kar, S., Baumann, W. T., Pau,l M. R. and Tyson, J. J., 2009. 
Exploring the roles of noise in the eukaryotic cell cycle. 
\textit{Proceedings of the National Academy of Sciences of USA}, \textbf{106}: 6471--6476.

\bibitem{DNAComputing1} 
Soloveichik, D., Seeling, G., Winfree, E., 2010.
DNA as a universal substrate for chemical kinetics.
\textit{Proceedings of the National Academy of Sciences}, \textbf{107}(\textbf{12}): 5393--5398.

\bibitem{DNAComputing2} 
Soloveichik, D., Cook, M., Winfree, E., Bruck, J., 2008.
Computation with finite stochastic chemical reaction networks.
\textit{Natural Computing}, \textbf{7}(\textbf{4}): 615--633.

\bibitem{MeRobust}  T. Plesa, G. B. Stan, T. E. Ouldridge, and W. Bae., 2020.
Quasi-robust control of biochemical reaction networks via stochastic morphing. 
Available as https://arxiv.org/abs/1908.10779.

\bibitem{Me1}  Plesa, T., Vejchodsk\'{y}, T., and Erban, R., 2016. 
Chemical reaction systems with a homoclinic bifurcation: An inverse problem. 
\textit{Journal of Mathematical Chemistry}, \textbf{54}(\textbf{10}): 1884--1915.

\bibitem{Me3} Plesa, T., and Zygalakis, K. C., Anderson, D. F., and Erban, R., 2018. Noise control for molecular computing.
\textit{Journal of the Royal Society Interface}, \textbf{15}(\textbf{144}): 20180199.

\bibitem{DNAComputing3} 
Srinivas, N., Parkin, J., Seeling, G., Winfree, E., Soloveichik, D., 2017.
Enzyme-free nucleic acid dynamical systems.
\textit{Science}, \textbf{358}, eaal2052.

\bibitem{Gillespietri} Gillespie, D. 
\textit{Markov processes: An introduction for physical scientists}. 
Academic Press, Inc., Harcourt Brace Jovanowich, 1992.

\bibitem{Experiment5} Zhang, D. Y., Winfree, E., 2009.
 Control of DNA strand displacement kinetics using toehold exchange. 
\textit{Journal of the American Chemical Society}, \textbf{131}: 17303--17314. 

\bibitem{Schlogl} Schl\"{o}gl, F., 1972. Chemical reaction models for nonequilibrium
phase transition. \emph{Z. Physik.}, \textbf{253}(\textbf{2}): 147--161.

\bibitem{Brusselator} Prigogine, I., and Lefever, R., 1968. 
Symmetry breaking instabilities in dissipative systems II.
 \textit{Journal of Chemical Physics}, \textbf{48}(\textbf{4}): 1695--1700.

\bibitem{Schnakenberg} Schnakenberg, J., 1979. 
Simple chemical reaction systems with limit cycle behaviour.
\textit{Journal of Theoretical Biology}, \textbf{81}(\textbf{3}): 389--400.

\bibitem{Me2}  Plesa, T., Vejchodsk\'{y}, T., and Erban, R, 2017. 
Test models for statistical inference: Two-dimensional reaction systems displaying 
limit cycle bifurcations and bistability, 2017. 
\textit{Stochastic Dynamical Systems, Multiscale Modeling, Asymptotics and 
Numerical Methods for Computational Cellular Biology}. 

\bibitem{RadekSNIPER} Erban, R., Chapman, S. J., Kevrekidis, I. and Vejchodsky, T., 2009. 
Analysis of a stochastic chemical system close to a SNIPER bifurcation of its mean-field model. 
\textit{SIAM Journal on Applied Mathematics}, \textbf{70}(\textbf{3}): 984--1016.

\bibitem{Radek2} 
Cao, Y., and Erban, R., 2014.
Stochastic Turing patterns: analysis of compartment-based approaches.
\textit{Bulletin of Mathematical Biology}, \textbf{76}(\textbf{12}): 3051--3069.

\bibitem{Radek3} Li, F., Chen, M., Erban, R., Cao, Y., 2018. 
Reaction time for trimolecular reactions in compartment-based reaction-diffusion models.
 \textit{Journal of Chemical Physics}, \textbf{148}, 204108.

\bibitem{Highmol1} Tyson, J. J, 1973. 
Some further studies of nonlinear oscillations in chemical systems. 
\textit{The Journal of Chemical Physics}, \textbf{58}, 3919.

\bibitem{Highmol2} Cook, G. B., Gray, P., Knapp, D. G., Scott, S. K., 1989. 
Bimolecular routes to cubic autocatalysis.
\textit{The Journal of Chemical Physics}, \textbf{93}: 2749--2755.

\bibitem{UNI3} Wilhelm, T., 2000. Chemical systems consisting only of elementary steps - a paradigma for nonlinear behavior. 
\textit{Journal of Mathematical Chemistry}, \textbf{27}: 71--88.

\bibitem{UNI1} Kerner, E. N., 1981. Universal formats for nonlinear ordinary differential systems. 
\textit{Journal of Mathematical Physics}, \textbf{22}: 1366--1371.

\bibitem{Kowalski} Kowalski, K., 1993. Universal formats for nonlinear dynamical systems. 
\textit{Chem. Phys. Lett.}, \textbf{209}: 167--170.

\bibitem{Vesicles2} Weitz, M., Kim, J., Kapsner, K., 
Winfree, E., Franco, E., Simmel, F. C., 2014.
Diversity in the dynamical behaviour of a compartmentalized programmable biochemical oscillator.
\textit{Nature Chemistry}, \textbf{6}: 295--302.

\bibitem{Vesicles3} Genot, A. J., Baccouche, A., Sieskind, R., Aubert-Kato, N., Bredeche, N., Bartolo, J. F., et al., 2016.
High-resolution mapping of bifurcations in nonlinear biochemical circuits.
\textit{Nature Chemistry}, 10.1038/nchem.2544.

\bibitem{GillespieDerivation} Gillespie, D. T., 1992. A rigorous derivation of the chemical master equation.
 \textit{Physica A: Statistical Mechanics and its Applications}, \textbf{188}(1): 404--425.

\bibitem{Janssen} Janssen, J., 1989. The elimination of fast variables in complex chemical reactions. II. Mesoscopic level (reducible case). 
\textit{Journal of Statistical Physics}, \textbf{57}: 171--185.

\bibitem{GrimaQSA} Thomas, P., Straube, A. V., and Grima, R., 2011. Communication: limitations of the
 stochastic quasi-steady-state approximation in open biochemical reaction networks. \textit{The Journal of Chemical Physics}, \textbf{135}(\textbf{18}): 181103.

\bibitem{KimQSA} Kim, J., Josic, K., and Bennett, M., 2014. The validity of quasi-steady-state approximations
 in discrete stochastic simulations. \textit{Biophysical Journal}, \textbf{107}: 783--793.

\bibitem{AgarwalQSA} Agarwal, A., Adams, R., Castellani, G. C., and Shouval, H. Z., 2012. On the precision of quasi steady state assumptions in stochastic dynamics. \textit{The Journal of Chemical Physics}, \textbf{137}: 044105.

% Radek's book.
\bibitem{RadekBook} Erban, R., Chapman, J. 
\emph{Stochastic Modelling of Reaction-Diffusion Processes}. 
Cambridge Texts in Applied Mathematics, Cambridge University Press, 2019. 

\bibitem{VanKampen} Van Kampen, N. G. \emph{Stochastic processes in physics and chemistry}. 
Elsevier, 2007.

\bibitem{PavliotisMultiscale} Pavliotis, G. A., Stuart, A. M. \emph{Multiscale methods: Averaging and homogenization}. 
Springer, New York, 2008.

\end{thebibliography}
\end{document}